\documentclass[11pt,a4paper]{article}
\usepackage{jheppub}
\pdfoutput=1

\usepackage{amsfonts}
\usepackage{amssymb}
\usepackage{amsmath}
\usepackage{color}
\usepackage[small,bf]{caption}
\usepackage[usenames,dvipsnames]{xcolor}
\usepackage{subfigure}
\usepackage{verbatim}
\usepackage{dsfont}
\usepackage[utf8]{inputenc}
\usepackage{tikz}
\usepackage{braket}
\usepackage{lscape}
\usepackage{slashed}
\usepackage{booktabs}
\usetikzlibrary{shapes,arrows}
\usepackage{mathtools}
\usepackage{graphicx}
\usepackage{fontawesome}
\usepackage{enumitem}
\usepackage{multirow}
\hypersetup{colorlinks=true,allcolors=[rgb]{1,0.56,0}}

\def\eps{\epsilon}
\def\1{\mathbf{1}}
\def\3{\mathbf{3}}
\def\2{\mathbf{2}}

\def\e{\varepsilon}

\allowdisplaybreaks[1]

\numberwithin{equation}{section}

\numberwithin{equation}{section}

\begin{document}\unitlength = 1mm

\title{A Statistical Analysis of the COHERENT Data and Applications to New Physics}

\author{Peter B.~Denton,}
\author{Julia Gehrlein}
\affiliation{High Energy Theory Group, Physics Department, Brookhaven National Laboratory, Upton, NY 11973, USA}
\emailAdd{pdenton@bnl.gov}
\emailAdd{jgehrlein@bnl.gov}
\date{\today}

\abstract{
The observation of coherent elastic neutrino nucleus scattering (CE$\nu$NS) by the COHERENT collaboration in 2017 has opened a new window to both test Standard Model predictions at relatively low energies and probe new physics scenarios. 
Our investigations show, however, that a careful treatment of the statistical methods used to analyze the data is essential to derive correct constraints and bounds on new physics parameters. In this manuscript we perform a detailed analysis of the publicly available COHERENT CsI data making use of all available background data. We point out that  Wilks' theorem is not fulfilled in general and a calculation of the confidence regions via Monte Carlo simulations following a Feldman-Cousins procedure is necessary.
As an example for the necessity of this approach to test new physics scenarios we quantify the allowed ranges for several scenarios with neutrino non-standard interactions. 
Furthermore, we provide accompanying code to enable an easy implementation of other new physics scenarios as well as data files of our results.
{\begin{center}\href{https://github.com/JuliaGehrlein/7stats}{\large\faGithub}\end{center}}}

\maketitle

\section{Introduction}
Coherent elastic neutrino nucleus scattering (CE$\nu$NS) is a Standard Model (SM) process that was predicted in 1974 \cite{Freedman:1973yd, Drukier:1983gj} and it describes the process when a neutrino scatters simultaneously and coherently with all of the nucleons in a nuclear target. The individual nucleonic amplitudes sum up coherently
which enhances the cross section by the number of nucleons squared.
Despite its large cross section, observation of this process is challenging due to the small nuclear recoil energies involved. Nevertheless due to the experimental advances in the last decade in detecting low recoil energies, CE$\nu$NS 
has been observed for the first time in 2017 by the COHERENT collaboration \cite{Akimov:2017ade}.
Since then a large interest in constraining standard and beyond the Standard Model (BSM) physics with this measurement has emerged.
The abilities of CE$\nu$NS to probe SM parameters at low momentum transfer, test new neutrino interactions, search for sterile neutrinos, its implications for supernova physics, as well as for dark matter searches, constraints on neutrino magnetic moments and nuclear physics show the far reaching influence of this process in many different areas of particle physics \cite{Dent:2016wcr,Coloma:2017egw,Ge:2017mcq,Coloma:2017ncl,Liao:2017uzy,Dent:2017mpr, Lindner:2016wff,Abdullah:2018ykz,Shoemaker:2017lzs,Kosmas:2017tsq,Farzan:2018gtr,Brdar:2018qqj,Datta:2018xty,Kosmas:2017zbh,Blanco:2019vyp,Denton:2018xmq, Canas:2019fjw, Chang:2020jwl, Flores:2020lji, Abdullah:2020iiv, Miranda:2020tif, Li:2020lba,Bowen:2020unj, Hurtado:2020vlj,Billard:2018jnl,AristizabalSierra:2018eqm,Bednyakov:2018mjd,Akhmedov:2018wlf,Cadeddu:2018dux,Heeck:2018nzc,Altmannshofer:2018xyo,AristizabalSierra:2019zmy,Cadeddu:2017etk,Ciuffoli:2018qem,Huang:2019ene,Papoulias:2019lfi,Dutta:2020che,Dutta:2019eml,Bednyakov:2019dbl,Miranda:2019wdy,AristizabalSierra:2019ufd,Canas:2018rng,Cadeddu:2018izq,Dutta:2019nbn,Miranda:2019skf,Papoulias:2019txv,Khan:2019cvi,Cadeddu:2019eta,Giunti:2019xpr,Han:2019zkz,AristizabalSierra:2019ykk,Papoulias:2019xaw,Arcadi:2019uif, Miranda:2020zji,Sahu:2020kwh,Pattavina:2020cqc,Han:2020pff,Foguel:2020fjx,Cadeddu:2020lky, Sadhukhan:2020etu,Coloma:2020nhf,Dutta:2020vop,1805471,Hoferichter:2020osn,Skiba:2020msb,Dutta:2020enk,1810387,Cadeddu:2020nbr,Tomalak:2020zfh,Suliga:2020jfa}. For this reason a careful statistical analysis is crucial to derive robust and reliable constraints on SM and BSM physics scenarios from CE$\nu$NS.

In this manuscript we will revisit and go beyond previous analyses to derive
statistically robust constraints on SM and BSM parameters using the publicly available data from the COHERENT CsI observation \cite{Akimov:2018vzs}. We will consider the timing and energy information provided in the data release. We find that the test statistic for the CsI data is not distributed according to a $\chi^2$ distribution with the number of degrees of freedom given by the number of energy and timing bins making a calculation of the $p$ value via Monte Carlo (MC) simulations  with the Feldman-Cousins \cite{Feldman:1997qc} approach necessary.
As an example for a new physics scenario we quantify the allowed ranges for neutrino non-standard interactions (NSI). For the first time in the literature we determine the allowed regions for all five relevant NSI parameters using MC simulations. 

Our analysis differs from previous analyses not only by the statistical approach but also in the treatment of the background and signal. Instead of considering only the ``anti-coincidence beam on" data set as background we additionally consider the beam off data as background to enhance the background statistics. Furthermore, we
address the question of binning for the new physics example of NSI.
Our analysis is implemented in a code which can be downloaded from \cite{7stats}. We also provide our results in the form of data files which can be downloaded from the same source. 

This manuscript is organized as follows: in section \ref{sec:overview}
we will give an overview of the COHERENT experiment and the CE$\nu$NS process, in section \ref{sec:num} we describe the calculation of the signal and background events at the COHERENT CsI detector. Section \ref{sec:analysis} is devoted to the analysis of the CsI data, in section \ref{sec:nsi} we use the data to derive constraints on the allowed region for NSI parameters in different scenarios, and finally we summarize and conclude in section \ref{sec:conc}.

\section{Overview of the Process}
\label{sec:overview}
\subsection{CE\texorpdfstring{$\nu$}{v}NS in the Standard Model}
CE$\nu$NS is a neutral current process which takes place for neutrino energies below about 50 MeV. In the SM it is mediated by the $Z$ boson.
The cross section for CE$\nu$NS for a neutrino of flavor $\alpha$ is given by \cite{Freedman:1973yd}
\begin{align}
 \frac{\text{d}\sigma_\alpha}{\text{d}E_R}=\frac{G_F^2}{2\pi}\frac{Q_{w\alpha}^2}{4}F^2(2 M E_R)M \left(
 2 -\frac{M E_R}{E_\nu^2}-2 \frac{E_R}{E_\nu}+\frac{E_R^2}{E_\nu^2}\right)\,,
\end{align}
where $G_F$ is the Fermi constant, $E_R$ is the nuclear recoil energy, $F(Q^2)$ is the nuclear form factor, $M$ is the mass of the target nucleus, and $E_\nu$ is the incident neutrino energy. In the kinematic term on the right the last two terms in are suppressed by $E_R/E_\nu$ in comparison to the first terms. The weak charge is given by 
\begin{align}
 \frac{Q_{w\alpha}^2}{4}&=\left(Zg_p^V +Ng^V_n\right)^2\,.
 \label{eq:Qw}
\end{align}
Here, $N$ and $Z$ are the number of neutrons and protons in the target nucleus.
The SM couplings of the Z boson to protons and neutrons at low energies are given by 
\cite{Barranco:2005yy}
\begin{align}
g_p^V&=\rho_{\nu N}^{NC}\left(\frac{1}{2}-2\kappa \hat{s}_Z^2\right)+2\lambda_{uL}+2\lambda_{uR}+\lambda_{dL}+\lambda_{dR}~,\nonumber\\
g_n^V&=-\frac{1}{2}\rho_{\nu N}^{NC}+\lambda_{uL}+\lambda_{uR}+2\lambda_{dL}+2\lambda_{dR}
\end{align}
with $\rho_{\nu N}^{NC}= 1.0082,~ \hat{s}_Z^2= \sin^2\theta_W= 0.23129,~\kappa= 0.9972,~ \lambda_{uL}=-0.0031,~\lambda_{dL}=-0.0025,$\\$\lambda_{dR}= 2\lambda_{uR}= 7.5\cdot 10^{-5}$. The contribution from the coupling to strange sea quarks in the nucleus is negligible for the process considered here \cite{ Beacom:2002hs}.
The nuclear form factor $F(Q^2)$ depends on the nuclear density distribution and is related to the physical size of the nucleus. It accounts for loss of coherency at higher values of momentum transfer, for small momentum transfer $F\sim 1$. 
In agreement with the official COHERENT analysis \cite{Akimov:2017ade} we choose the Klein Nystrand parametrization \cite{Klein:1999qj} for the form factor which is given as
\begin{align}
F(Q^2)=\frac{4 \pi \rho}{A Q^3}(\sin(Q R_a)-Q R_a \cos(Q R_a))\frac{1}{1+a^2 Q^2}\,,
\end{align}
with $a=0.7$ fm, $R_a=1.2A^{1/3} $ fm, and $\rho=3 A/(4\pi R_a^3)$.
Using a different form factor has only a small effect on the number of signal events; nonetheless we will account for the form factor uncertainty in the analysis in section \ref{sec:nsi} by including a systematic uncertainty on the normalization of the signal.

The expected rate of CE$\nu$NS events depends on the specifications of the
detector considered and the timing and energy structure of the neutrino source. In the following we will focus on the CsI detector of the COHERENT
experiment.
The COHERENT experiment is located at the Spallation Neutrino Source (SNS) at Oak Ridge National Laboratory. The pulsed source produces $\pi^+$ and $\pi^-$ in proton-nucleus collisions in a mercury target. The $\pi^-$ are absorbed by nuclei before they can decay, the $\pi^+$ lose energy as they propagate and eventually decay at rest into $\pi^+\to \mu^++\nu_\mu$, followed by the muon decay $\mu^+\to e^++\overline{\nu}_\mu+\nu_e$. As the muon lifetime is much longer than that of the pion the monochromatic $\nu_\mu$ component (at $E_{\nu_\mu}= (m_\pi^2-m_\mu^2)/(2m_\pi)\approx 29.7 $ MeV) is referred to as prompt flux, while the delayed neutrino flux from muon decay $\nu_e$ and $\overline{\nu}_\mu$ has a continuous energy spectrum up to $E_{\nu_e,\overline{\nu}_\mu}< m_\mu/2\approx 52.8$ MeV. From simple decay kinematics the 
incoming neutrino flux for the COHERENT experiment are given as,
\begin{align}
f_{\nu_\mu}&=\delta\left(E_\nu-\frac{m_\pi^2-m_\mu^2}{2m_\pi}\right)\,,\nonumber\\
f_{\overline{\nu}_\mu}&=\frac{64}{m_\mu}\left[\left(\frac{E_\nu}{m_\mu}\right)^2\left(\frac{3}{4}-\frac{E_\nu}{m_\mu}\right)\right]\,,\nonumber\\
f_{\nu_e}&=\frac{192}{m_\mu}\left[\left(\frac{E_\nu}{m_\mu}\right)^2\left(\frac{1}{2}-\frac{E_\nu}{m_\mu}\right)\right]\,,
\label{eq:per flavor flux}
\end{align}
for neutrino energy $E_\nu\in [0,m_\mu/2]$ with $m_\mu$ and $m_\pi$ the muon and pion mass.
This simple approximation is in excellent agreement with the more realistic simulation of the flux from the SNS \cite{Akimov:2017ade}.

The different energy and timing structure of the different flavors leads to an increased distinguishing power of flavor dependent new physics like NSI \cite{Denton:2018xmq,Coloma:2019mbs}.
For this reason we will analyze in the following the timing and energy distribution of the number of events.

To obtain the flux at the detector at distance $\ell$ these expressions need to be multiplied by the geometric factor $1/(4\pi\ell^2)$ and the number of neutrinos produced by the proton collisions on target during the running time. For the CsI detector $\ell=19.3$ m and $f_{\nu/p}=0.08$ neutrinos per proton collision at SNS are produced; uncertainties in $f_{\nu/p}$ are included in the normalization systematic. The released data uses a running time of 308.1 days during which time $N_{\rm POT}=1.76\cdot 10^{23}$ protons on target were accumulated.
The expected number of CE$\nu$NS events per nuclear recoil energy can by calculated with
\begin{align}
\label{eq:dnder}
\frac{\text{d}N}{\text{d}E_R}=\frac{N_{target}N_{\rm POT} f_{\nu/p}}{4\pi\ell^2} \int \text{d}E_\nu f_\alpha(E_\nu)\frac{\text{d}\sigma_\alpha (E_\nu)}{\text{d}E_R}\,,
\end{align}
where $N_{target}$ is the number of target nuclei.
The mass of the CsI detector is 14.6 kg. The atomic numbers of the Cs and I nucleus are similar ($A_I=127,~ Z_I=53$ and $A_{Cs}=133,~Z_{Cs}=55$) nevertheless we calculate the individual cross sections separately and weight the number of events according to the nuclear masses.

\subsection{NSI Review}
In BSM theories other neutral particles can contribute to CE$\nu$NS. The effect of new heavy mediators can be parametrized in an effective field theory approach when the mediator mass exceeds the energy transfer $Q$. For NSI affecting CE$\nu$NS at COHERENT this is the case for mediator masses above $\sim$100 MeV \cite{Denton:2018xmq}. The framework for such NSI is given by \cite{Wolfenstein:1977ue}.
In the NSI framework neutrinos experience a new, possibly flavor changing, interaction governed by an addition to the Lagrangian,
 \begin{align}
 \label{eq:Lnsi}
 \mathcal{L}_{\text{NSI}}= -2\sqrt{2}G_F\sum_{\alpha,\beta,f}\epsilon ^{f,V}_{\alpha\beta}(\overline{\nu}_\alpha\gamma_\mu P_L\nu_\beta)(\overline{f}\gamma^\mu P f)\,,
\end{align}
where $\alpha$ and $\beta$ refer to the flavor indices of the neutrinos, $P=P_L,~P_R$, and $f$ stands for SM charged fermion typically $e$, $u$, or $d$. In this notation, $\epsilon^f_{\alpha\beta}$ provides an effective field theory parametrization of the strength of the new interaction with respect to the Fermi constant.
That is, $\epsilon^f_{\alpha\beta}\sim \mathcal{O}(G_X/G_F)$ with the new physics effective coupling $G_X$.

Many interesting UV complete NSI models leading to $|\eps_{\alpha\beta}^{f,V}|\sim\mathcal O(0.1)$ or larger have been developed in recent years \cite{Forero:2016ghr,Denton:2018dqq,Dey:2018yht,Babu:2017olk,Farzan:2016wym,Farzan:2015hkd,Farzan:2015doa,Babu:2019mfe}.
NSI provides useful means of connecting new physics processes involving neutrinos to oscillation physics.
In fact, there have been several hints in oscillation data of new physics that could be explained by NSIs $|\eps_{\alpha\beta}^{f,V}|\sim0.01-0.1$ \cite{Girardi:2014kca,Liao:2016reh,Denton:2018dqq,Capozzi:2019iqn,Denton:2020uda}.
For an overview of the breadth of NSI physics, see \cite{Dev:2019anc}.

In general an axial coupling can be present in neutral current NSI, but the effect is negligible in coherent elastic scattering of neutrinos with heavy nuclei \cite{Barranco:2005yy} and becomes only important for light targets like Na as the relative contribution depends on the inverse of the number of nucleons assuming comparable axial and vector couplings. As the constraints for scalar interactions are qualitatively similar to the constraints on vector interactions we will focus only on vector interactions. 
The weak charge from eq.~\eqref{eq:Qw} is now replaced by,
\begin{align}
\label{eq:QWnsi}
 \frac{Q_{w\alpha}^2}{4}={}&\left[Z(g_p^V+ 2\epsilon^{u,V}_{\alpha \alpha}+\epsilon^{d,V}_{\alpha\alpha}) +N(g^V_n+\epsilon^{u,V}_{\alpha\alpha}+ 2\epsilon^{d,V}_{\alpha \alpha})\right]^2\nonumber\\&+\sum_{\beta\neq\alpha}\left[Z(2\epsilon^{u,V}_{\alpha\beta}+\epsilon^{d,V}_{\alpha\beta}) +N(\epsilon^{u,V}_{\alpha\beta}+ 2\epsilon^{d,V}_{\alpha\beta})\right]^2\,.
\end{align}
In principle if the mediator for the interaction is of a similar mass as the energy scale of the experiment, it can provide an energy dependent effect to eq.~\eqref{eq:QWnsi}.
A light mediator can also affect the value of the form factor and the values of $g_p^V, g_n^V$ via running effects.
We focus only on the heavy mediator case here, where heavy means $\gtrsim$100 MeV.
We also assume that the new physics is dominantly on the neutrino side and that the quark--mediator coupling is smaller than the neutrino--mediator coupling.
Due to a degeneracy in the cross section there are two distinct values of $\epsilon_{\alpha\alpha}$ which lead to the same cross section, hence we expect two exact degenerate minima in the test statistic when the term inside the square brackets on the first line of eq.~\eqref{eq:QWnsi} changes sign.

As the CE$\nu$NS process with only one target nuclei is not sensitive to the difference between up and down quarks we consider $\epsilon^{u,V}_{\alpha\beta}=\epsilon^{d,V}_{\alpha\beta}$ (our results can be easily translated to other up to down quark ratios) as in \cite{Denton:2018xmq}. As COHERENT is not sensitive to $\epsilon_{\tau\tau}$
we are left with 5 parameters to constrain ($\epsilon_{ee}^V,~\epsilon_{\mu\mu}^V,~\epsilon_{e\mu}^V,~\epsilon_{e\tau}^V,~\epsilon_{\mu\tau}^V$).
The remaining NSI parameter under these simplifications, $\eps^V_{\tau\tau}$, can be connected to these parameters via the tight constraint on $\eps^V_{\tau\tau}-\eps^V_{\mu\mu}$ from oscillations.

\section{CE\texorpdfstring{$\boldsymbol{\nu}$}{v}NS at COHERENT CsI}
\label{sec:num}
In order to obtain the CE$\nu$NS signal events at the COHERENT CsI detector the following procedure needs to be implemented. 

Eq.~\eqref{eq:dnder} provides the number of events per nuclear recoil energy.
However, the COHERENT CsI detector does not directly measure nuclear recoil energy; instead it records the number of photoelectrons (PE) produced by an event. Then one can relate $E_R$ to the number of PE by the quenching factor $Q$ and the light yield $Y$.
The quenching factor accounts for the fact that when the nucleus recoils its energy is dissipated through a combination of scintillation (ionization of the material) and secondary nuclear recoils (heat).
The characteristic signal of a nuclear recoil are secondary recoils, however 
their measurable signal is much smaller than that of electron recoils. The ratio between the light yields from a nuclear and an electron recoil of the same energy is the quenching factor $Q$. The light yield parametrizes the amount of electron recoil energy which is converted into PE. The light yield is provided in the data release as $Y=13.348\pm0.019$ PE/keVee where keVee is the electron recoil energy in keV \cite{Akimov:2018vzs}. The measurement of the quenching factor is fairly complicated and a number of measurements of $Q$ by several collaborations have been performed. The agreement of the measurements, however, is quite poor. A recent measurement \cite{Collar:2019ihs} which claimed smaller error bars could not be confirmed by the COHERENT collaboration.
For this reason we follow the recommendation of the collaboration \cite{Konovalov2019} and use the quenching factor quoted in the data release which is independent on the nuclear recoil energy and whose error bars encompass the lowest and highest measurement of the quenching factor in the considered region for nuclear recoils. This gives $Q=8.78\pm1.66\%$ \cite{Akimov:2018vzs}.
The uncertainty on both light yield and quenching factor only affect the normalization of the signal and are in secs.~\ref{sec:analysis}, \ref{sec:nsi} taken into account as pull terms on the signal.

It is important to note that the number of PE produced in a given interaction follows a Poisson distribution and an additional smearing must be accounted for to relate the expected number of PE, PE$_{\rm raw}$ to the observed number, PE$_{\rm true}$.
The relation between the number of events in the raw PE bins before smearing and the number of events in the true PE bins after smearing is given by,
\begin{align}
\label{eq:smear}
 N(\text{PE}_{\text{true}})=\sum_{\text{PE}_{\text{raw}}}N(\text{PE}_{\text{raw}}) P(\text{PE}_{\text{true}},\text{PE}_{\text{raw}})\,,
\end{align}
where the Poisson distribution is
\begin{align}
\label{eq:poiss}
 P(k,\lambda)=\text{e}^{-\lambda} \frac{\lambda^{k}}{k!}\,.
\end{align}
The effect of smearing depends on the number of true energy bins considered for the analysis.
If only one energy bin is considered (that is, energy information is not included in the analysis) then smearing has no effect\footnote{The inclusion of cuts leads to a slight effect here, but it turns out to be negligible for the PE cuts typically considered.}.
For more energy bins, in particular for all 12 available energy bins, we have found that smearing needs to be included.

Smearing also affects the background events. 
At COHERENT there are two sources of backgrounds\footnote{Another background source is neutrino-induced neutrons (NINs) that originate in the shielding surrounding the detector. However it has been shown that NINs is negligible at the location of COHERENT \cite{Akimov:2017ade} and is hence ignored in the following.}. The neutron background comes from the neutrons produced in the beam; their arrival time is the same as the prompt neutrino flux. 
The steady state background is coming from either cosmic rays or their by-products entering the detector and their arrival time is not related to the beam time. 

\begin{figure}
 \centering
 \includegraphics[width=0.48\linewidth]{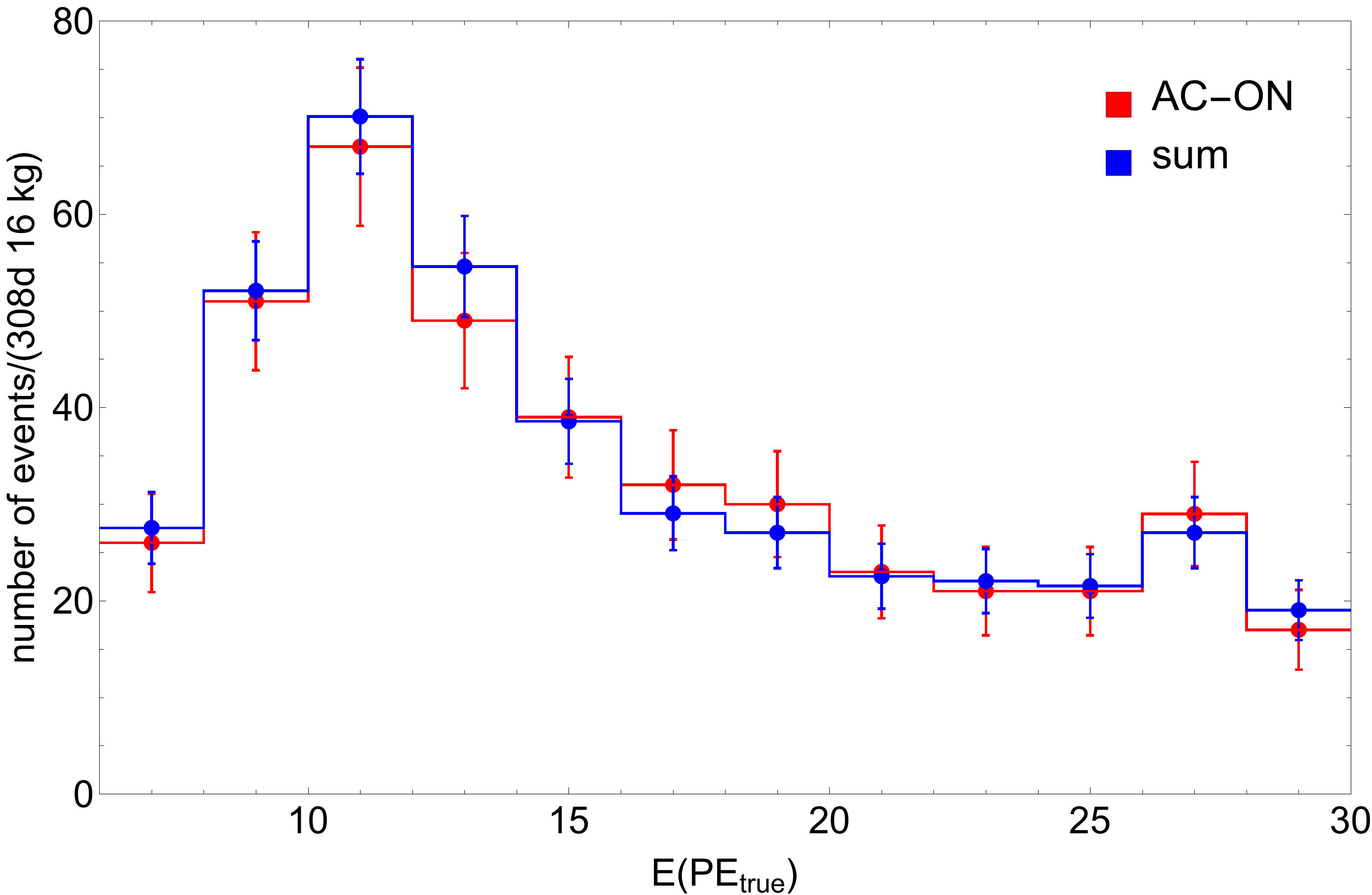}
 \includegraphics[width=0.48\linewidth]{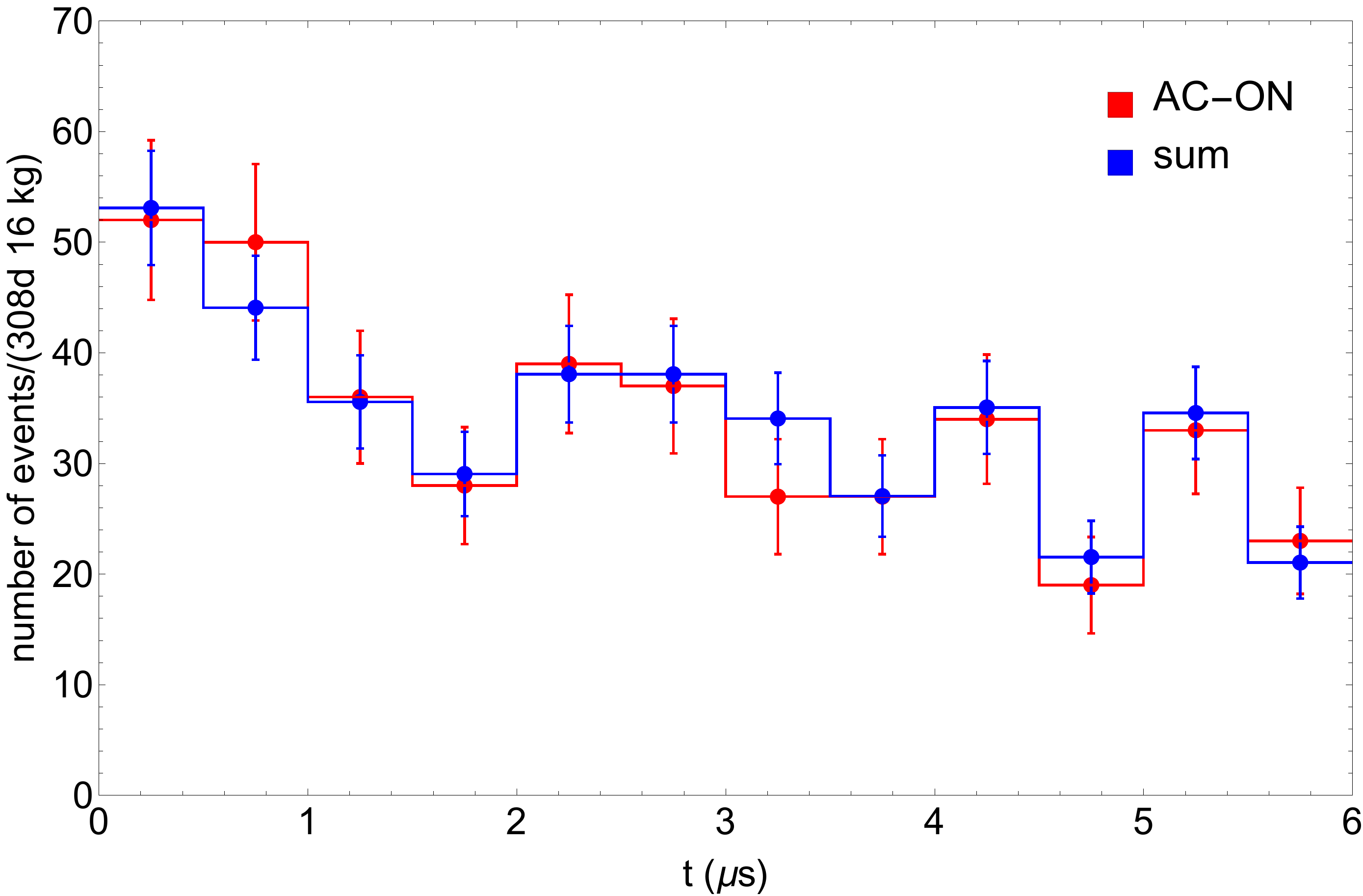}
 \caption{Comparison of the AC-ON background and the rescaled sum of all background templates for the 1D distributions of energy and time. The error bars show the statistical uncertainties where for the sum of all background data the relative uncertainty of the unweighted sum of backgrounds has been used.}
 \label{fig:1dcomp}
\end{figure}

The COHERENT collaboration released the results of a simulation of the energy (in terms of raw PE bins) and timing distribution of the neutron background. 
Additionally the collaboration released four data files. The coincidence beam on data (C-ON) which is the ``signal" data, the coincidence beam off data (C-OFF) and the anti-coincidence beam on and beam off data (AC-ON and AC-OFF) files as the ``background" data. 
The difference between coincidence and anti-coincidence is determined by two different timing windows where the coincidence window is when the SNS beam arrives (C-OFF is when the beam would arrive if it were on), contributions from the SNS beam are only expected in the coincidence window \cite{Akimov:2017ade,Rich:2017lzd}.
As the background is expected to be uncorrelated between beam on and off and in particular not related to the coincidence or anti-coincidence regions all three templates should be used as background to increase the background statistics and hence lower the uncertainty on the background normalization. This approach differs from all previous analyses which only used the AC-ON data to estimate the background. All three backgrounds are compatible, as expected (see fig.~\ref{fig:1dcomp} for a comparison of the 1D projections of the AC-ON data and the rescaled sum of all three backgrounds).
We then perform a weighted sum (the beam off data includes 153.5 days of exposure compared to 308.1 days of exposure when the beam was on \cite{Akimov:2017ade}) to account for the different exposures and refer to this as the steady state background. 
The steady state background is the largest contribution to the number of events. In fact the number of background events is more than twice as large as the number of signal events making a correct treatment of the background crucial to derive constraints on the signal.

We have confirmed that the timing and energy of the background are uncorrelated, consistent with previous analyses by COHERENT \cite{Akimov:2017ade,Rich:2017lzd}.
For this reason it is possible 
to factorize the background data producing 1D projections in energy and time to reduce the effect of statistical fluctuations in the background model. 
The full 2D template can then be obtained by multiplying these two distributions.
We follow this approach after accounting for the cuts in energy and time which are $7\leq \text{PE}_{\text{true}}\leq 30$ and $t\leq6~\mu$s from the beginning of the pulse.

Smearing needs to be applied to the sum of signal and background events according to eq.~\eqref{eq:smear}.
However the steady state background is only provided as measured events in true PE bins and not as events in raw PE bins.
In order to circumvent this problem and to obtain self consistent results, one could attempt to invert the effect of smearing and efficiency on the background and infer the underlying functional dependence by considering various functional forms for the true background and then applying smearing and efficiency and comparing to the data.
After trying many functional forms we found, however, that none provided an acceptable fit to the data.

The factorization of the background data is only valid so long as the energy distribution doesn't depend on the time.
One could also attempt to reconstruct the entire 2D distribution spanning 144 bins and accounting for the detector efficiency and smearing and then fitting to the measured background data with 144 degrees of freedom.
We found this to be numerically unreliable due to the considerable fluctuations in the background data.
In addition, any attempt to simply parametrize the 2D background proved to be infeasible without a very large number of parameters in the fit which likely suffers from over fitting.
Finally, the features in the data, in particular the fact that the timing distribution is not flat, don't have a clearly motivated physics interpretation and simple functional fits do not provide a good fit to the timing distribution\footnote{We speculate that this could be due to pre-trace cuts applied to the data in the timing window before the coincidence or anti-coincidence timing windows, but are unable to verify this with publicly available data.}.
For this reason we use the method described above of projecting the data down to 1D to determine the timing and energy distributions, and then projecting back to the full 2D distribution.

Finally, the signal acceptance efficiency $C$ of the CsI detector need to be taken into account for the signal.
An analytic parametrization in terms of PE$_{\rm true}$ is provided in \cite{Akimov:2018vzs},
\begin{align}
 C(\text{PE}_{\text{true}})=\frac{a}{1+\text{exp}(-k (\text{PE}_{\text{true}}-x_0))}\Theta(\text{PE}_{\text{true}})\,,
\end{align}
with 
\begin{align}
\label{eq:eff}
a&= 0.6655^{+0.0212}_{-0.0384}\,,\nonumber\\
k&= 0.4942^{+0.0335}_{-0.0131}\,,\\
x_0&=10.8507^{+0.1838}_{-0.3995}\,,\nonumber
\end{align}
and 
\begin{align}
 \Theta(\text{PE}_{\text{true}})=\begin{cases}
0 \quad &\text{if $\text{PE}_{\text{true}}<5$}\\
0.5 \quad &\text{if $5\leq \text{PE}_{\text{true}}<6$}\\
1\quad &\text{if $\text{PE}_{\text{true}}\geq6$}
\end{cases}\,.
\end{align}

The integral over eq.~\eqref{eq:dnder} in raw PE space gives the energy distribution of the signal. The timing distribution of the signal has been provided by the collaboration \cite{Akimov:2018vzs}. Multiplying the two distributions gives the 2D timing-energy distribution. As the neutron background has been provided in terms of raw PE, it can be added to the signal such that smearing and efficiency can be applied to both distributions simultaneously.
Adding to this the steady state background distribution leads to the 2D distribution of expected events.
The uncertainties in the efficiency parameters in eq.~\eqref{eq:eff} have a small effect on our results; nonetheless, we include the impact of the uncertainty on $a$ into our normalization uncertainty.

\section{Analysis Methods}
\label{sec:analysis}
\subsection{Test statistic}
With the approach described in the previous section we calculate the $p$ value of the SM and confidence regions for new physics scenarios using a test statistic (TS).
After applying the cuts on the released data files the considered energy bins range from [7,30] $\text{PE}_{\text{true}}$ and timing between [0,6] $\mu$s. 

We use Poisson statistics to calculate the log likelihood ratio in the $i^{\rm th}$ timing and $j^{\rm th}$ energy bin and then sum over each bin and minimize over the nuisance parameters (pull terms)
\begin{align}
\label{eq:ts}
\chi^2=\min_{\alpha,\beta,\gamma}2\sum_{i,j}\left[T_{ij} - D_{ij} + D_{ij}\log\left(\frac{D_{ij}}{T_{ij}}\right)\right]+f_{\text{pull}}(\alpha,\sigma_\alpha)+f_{\text{pull}}(\beta,\sigma_\beta)+f_{\text{pull}}(\gamma,\sigma_\gamma)\,,
\end{align}
where $D_{ij}$ is for the number of events taken from the C-ON data set and $T_{ij}$ stands for the theoretically predicted number of events which is calculated as the sum of the steady state background events ($N^{bkg}_{ij}$) and the neutron background ($N^{neut}_{ij}$) plus signal ($N^{sig}_{ij}$) after their raw PE distribution has been smeared according to a Poisson distribution and the efficiency $C(\text{PE}_{\text{true}})$ has been applied 
\begin{align}
\label{eq:tij}
 T_{ij}=(1+\gamma)N^{bkg}_{ij}(\text{PE}_{\text{true}})+C(\text{PE}_{\text{true}})N_{ij}(\text{PE}_{\text{true}})\,,
\end{align}
where 
\begin{align}
 N_{ij}(\text{PE}_{\text{true}})=\sum_{\text{PE}_{\text{raw}}} s_{ij}(\text{PE}_{\text{raw}}) P(\text{PE}_{\text{true}},\text{PE}_{\text{raw}})\,,
 \label{eq:smearing}
\end{align}
with $P$ the Poisson distribution from eq.~\eqref{eq:poiss} multiplied by 
\begin{align}
s_{ij}(\text{PE}_{\text{raw}})=((1+\alpha)N^{sig}_{ij}(\text{PE}_{\text{raw}})+(1+\beta)N^{neut}_{ij}(\text{PE}_{\text{raw}}))\,.
\end{align} 
We include pull terms in the likelihood function to account for the uncertainty of the signal normalization $\alpha$, $\sigma_\alpha$, coming from the uncertainty on the quenching factor ($25\%$) (we prefer to include a larger uncertainty than quoted due to the disagreement between different measurements of $Q$ \cite{Akimov:2017ade,Rich:2017lzd,Scholz:2017ldm,Guo:2016ovb,Park:2002jr,Collar:2019ihs}), uncertainties on the
neutrino flux (10$\%$)\footnote{The uncertainty on the incoming neutrino flux provides the largest uncertainty after the quenching factor. A dedicated D$_2$O detector will be installed at SNS to improve the flux calibration and reduce the uncertainty in the future \cite{Tolstukhin2019}.}, light yield (0.14$\%$), signal acceptance (5$\%$), and form factor (5$\%$) \cite{Akimov:2017ade} added in quadrature. The uncertainty on the normalization of the neutron background $\beta$ is $\sigma_\beta=25\%$ \cite{Akimov:2017ade}. We calculate the uncertainty on the state steady background normalization $\gamma$ by using the square root of the average number of events per bin in the timing and energy distributions of the sum of the three (C-OFF, AC-ON, AC-OFF) unweighted background distributions and add them in quadrature.
A Gaussian prior is typically assumed for these pull terms.
However, when the true uncertainty on a given systematic is dominated by Poisson fluctuations in another measurement, is otherwise asymmetric, or when the pull term is getting pulled to $\sim\pm1$, a Gaussian pull term is typically no longer appropriate.
This means that for a small numbers of events, the likelihood function is usually skewed, resulting in asymmetric error intervals and pull distributions that are non-Gaussian \cite{Demortier}. In order to obtain the correct pull term distribution for the CsI data insights into the calculation of the uncertainties (for example the calculation of the uncertainty on the quenching factor or the MC calculation of the neutron background) is required. This information has not been published.
As an example, it can be already seen from eq.~\eqref{eq:eff} that the error on the normalization of the signal efficiency is asymmetric. Using a symmetric pull term can lead to unphysical results for this reason it is desirable that the collaboration provides the correct pull term parametrization.
We find that the impact of assuming a Gaussian pull terms can be large in the optimal binning configurations, see appendix \ref{sec:diff_stat} for further validation of this.

\subsection{Binning}
Due to the low statistics of the signal, it is desirable to reduce the fluctuations as much as possible while maintaining as much information as possible. 
In general one prefers to use an unbinned likelihood in this case however the COHERENT data has only been released in a binned form. Furthermore, it is preferable to use as many bins as possible.


To demonstrate the impact of different number of bins we present our results for various different binning configurations:
\begin{enumerate}[leftmargin=0.7in]
\item[\bf 1T,1E:] a counting experiment,
\item[\bf 2T,1E:] two timing bins [0,1] and [1,6] $\mu$s splitting the signal into the prompt and delayed components (see eq.~\eqref{eq:per flavor flux}),
\item[\bf 2T,4E:] in addition to the two timing bins in 2T,1E, we also bin the energy data into [7,18], [19,22], [23,26], and [27,30] PE, for eight total bins,
\item[\bf 12T,12E:] 12 timing bins with width 0.5 $\mu$s between 0-6 $\mu$s and 12 energy bins with width 2 PE from 7-30 PE for 144 total bins.
\end{enumerate}
Smearing (see eq.~\ref{eq:smearing}) is taken into account for the 2T,4E and 12T,12E configurations.

In general the boundaries of the bins need to be adapted according to the specific model to test. 
In principle, such an optimization should be done before data is collected for a given new physics scenario.

\section{Results}
\label{sec:nsi}
In this section we demonstrate our improved statistical approach to obtain constraints on SM and BSM parameters using MC estimations.
We first discuss the goodness of fit of the SM and then a collection of non-standard interactions scenarios.

\subsection[Goodness of fit $\&$ calculation of the $p$ value]
{Goodness of fit $\boldsymbol{\&}$ calculation of the $\boldsymbol{p}$ value}
The first thing we check is if the SM is a good fit to the data. To this aim we perform a MC simulation of COHERENT to calculate the $p$ value of the SM.
The procedure to obtain the $p$ value is as follow:
\begin{enumerate}
 \item For the physics scenario under consideration (for example assuming all NSI parameters to be non-zero simultaneously, or a certain bin configuration) we determine the best fit point by minimizing the TS over the pull terms and any NSI terms in eq.~\eqref{eq:ts} using the observed CsI data as $D_{ij}$.
 \item We then use the best fit NSI parameters to throw many pseudo experiments while randomly selecting the neutron background and signal normalization pull terms and fluctuating the statistics (signal plus backgrounds) in each bin.
 The resultant events are then used as ``data" $D_{ij}$ compared to the prediction from the best fit NSI parameters to calculate the TS with eq.~\eqref{eq:ts}.
\item We repeat this step $\mathcal{O}(10^4)$ to obtain the probability density function of the TS.
\item
For a certain point in parameter space (e.g.~$\eps_{ee}=0$) we calculate the TS in eq.~\eqref{eq:tij} again while minimizing over the other physics and nuisance parameters compared against the data.
The $p$ value is then the fraction of the TS's in the PDF from the previous step which are larger than this TS.
\end{enumerate}
In principle, under certain assumptions, this procedure can be skipped and Wilks' theorem can be applied.
Wilks' theorem allows for the easy extraction of the model preference and confidence intervals without the need for the often computationally expensive simulations described above.
In many applications, however, these conditions are not satisfied (or difficult to confirm) and a proper MC treatment is necessary.
In the following sections we will explicitly test these assumptions and find that Wilks' theorem is not satisfied in this context.

Turning now to our results.
We performed a MC simulation of COHERENT to calculate the $p$ value of the SM using the different bin configurations. In this case we assumed the SM to be the best fit point for the data (i.e. we assumed the SM to be reality) and followed the steps outlined above. Namely, we generated a PDF using the SM prediction as $N_{ij}^{sig}$ in eq.~\eqref{eq:tij}, took the signal and normalization pull terms from a normal distribution and applied Poisson fluctuations to the sum of signal and backgrounds according to eq.~\eqref{eq:tij} which we then used as $D_{ij}$ to calculate the TS with eq.~\eqref{eq:ts} to generate the PDF.

The results for the $p$ value are shown in tab.~\ref{tab:smpval}. We find that the SM is a good fit to the data independent of the number of bins used for the analysis.

\begin{table}
\centering
\caption{SM $p$ value for different bin configurations using the COHERENT CsI data.}
\begin{tabular}{c||c|c|c|c|c}
bin configuration&1T,1E&2T,1E&2T,2E&2T,4E&12T,12E\\\hline
$p$ value&0.68&0.34&0.85&0.48&0.64\\
\end{tabular}
\label{tab:smpval}
\end{table}

\subsection{NSI parameters one-at-a-time}
\label{sec:nsi_one_at_time}
In this section we will focus on the constraints for only one non-zero NSI parameter; the next section covers the case where all relevant NSI parameters are allowed to be non-zero at the same time.

To obtain the constraints on NSI parameters and determine their confidence levels we employ a Feldman-Cousins (FC) approach following these steps:
\begin{enumerate}
\item We start from our definition of the TS in eq.~\eqref{eq:ts} (see also \cite{Feldman:1997qc})
\begin{equation}
\chi^2(\vec\eps,\vec\alpha,\vec n)=2\sum_i\left[\mu_i(\vec\eps,\vec\alpha)-n_i+n_i\log\left(\frac{n_i}{\mu_i(\vec\eps,\vec\alpha)}\right)\right]+\sum_j\left(\frac{\alpha_j}{\sigma_{\alpha_j}}\right)^2
\end{equation}
with  theory prediction $\vec\mu(\vec\eps,\vec\alpha)$ that depends on physics parameters of interest $\vec\eps$ and nuisance parameters $\vec\alpha$. In our case $\vec\eps$ is either only one non-zero NSI parameter or all of them are allowed to be non-zero, and $\vec\alpha=(\alpha,\beta,\gamma)$ contains the normalization pull terms.
Again,  $n_i$ is the number of measured events in bin $i$.
\item We pick a value of new physics parameters $\vec\eps^*$  of interest and determine the best fit values of the nuisance parameters $\vec\alpha_{bf}(\vec\eps^*)$  using  the TS $\chi^2(\vec\eps^*,\vec\alpha,\vec n_{data})$ with $\vec\eps^*$ fixed and $\vec n_{data}$ corresponds to the real, measured data. This approach has been proposed in \cite{NOVA-doc-15884-v3} to ensure a proper coverage of all values of the nuisance parameters. As  we have covered for their most likely value all the less likely values will give us confidence belts that are contained in the one we calculated. 
\item We  simulate the experiment with a MC to obtain the event rate $\vec n_{MC}(\vec\eps^*,\vec\alpha_{bf}(\vec\eps^*))$ at the fixed value of $\vec\eps^*$ and $\vec\alpha_{bf}(\vec\eps^*)$ determined above. We then 
calculate $\Delta\chi^2=\min_{\vec\alpha}[\chi^2(\vec\eps^*,\vec\alpha,\vec n_{MC})]-\min_{\vec\eps,\vec\alpha}[\chi^2(\vec\eps,\vec\alpha,\vec n_{MC})]$, i.e. the difference between the TS where $\vec{\epsilon}$ is fixed to the value of interest to the TS where $\vec{\epsilon}$ is marginalized over.

\item Repeating the previous step many times we obtain a distribution of $\Delta \chi^2$ (PDF) which allows us to construct a confidence interval from the toy experiments at the desired confidence level. 
\item To obtain the allowed parameter range for the physics scenario of interest we repeat the previous three steps for many values of $\vec\eps^*$. 
\item  We then calculate for each  $\vec\eps^*$,  $\Delta\chi^2=\min_{\vec\alpha}[\chi^2(\vec\eps^*,\vec\alpha,\vec n_{data})]-\min_{\vec\eps,\vec\alpha}[\chi^2(\vec\eps,\vec\alpha,\vec n_{data})]$ using the observed data.
\item $\vec\eps^*$ is included in the confidence interval for the measurement if the $\Delta\chi^2$ of the data from the previous step is within the confidence interval for the $\Delta\chi^2$ from the MC simulations calculated in step 4.

\end{enumerate}
It should be noted that in general, one needs to check if using the best fit nuisance parameters in step 2 is correct.
To do this, repeat this exercise by varying $\vec\alpha_{bf}(\vec\eps^*)$ across  a range and check to see if any additional regions are included in the confidence interval (for more details see \cite{10.2307/2290928, NOVA-doc-15884-v3}).  
To be more precise one needs to search for parameters of interest  outside the confidence interval and values of  nuisance parameters not most favorable to the data, and verify that these points have confidence belts wholly within the one we have constructed for this value of the parameter of interest.

Furthermore, the FC procedure does not tell us if the best fit point is a good fit, meaning that the FC approach does not provide information on the $p$ value of the best fit point. As we will see in the following the best fit points we found are in the vicinity of the SM point. Since we have shown in the previous section that the SM is a good fit to the data also the BSM best fit point is expected to be a good fit as well.

\begin{figure}
 \centering
 \includegraphics[width=0.6\linewidth]{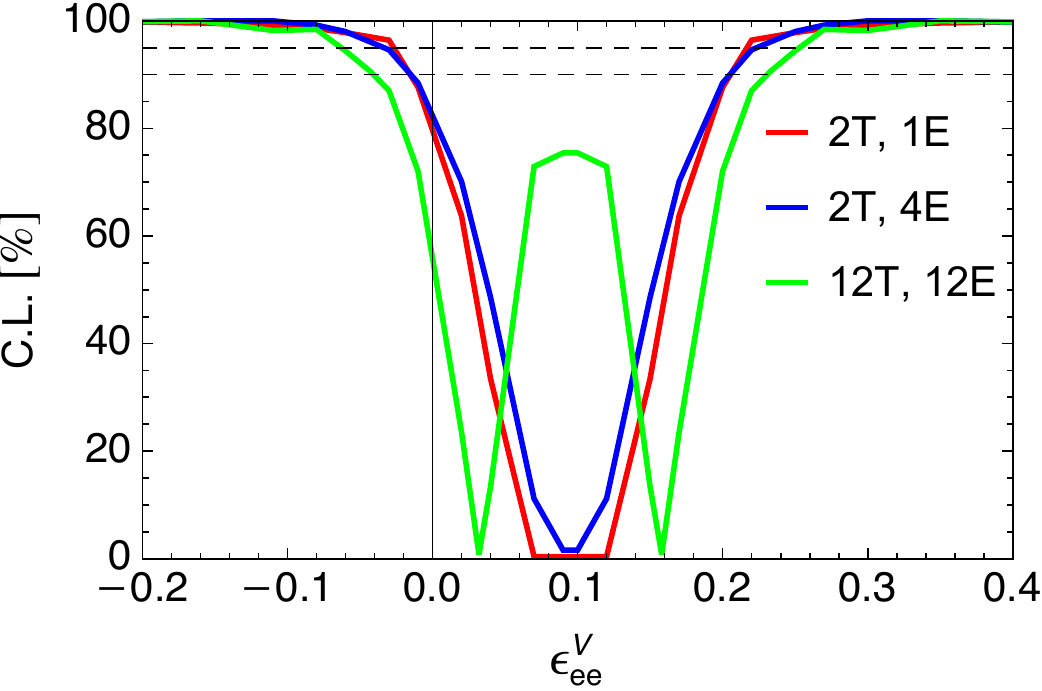}
 \caption{Results for $\eps_{ee}^V$ using different numbers of bins and Gaussian pull terms. We assume the remaining NSI parameters to be zero. }
 \label{fig:pvalnsi}
\end{figure}

Coming now to our results. The first question we address is the ideal number of bins. 
In fig.~\ref{fig:pvalnsi} we show the constraints on only non-zero $\eps_{ee}^V$ using 2T, 1E, 2T,4E, and 12T, 12E bins.
We see that only with the 12T, 12E bin configuration  $\eps_{ee}^V\approx 0.1$ can be ruled out, making this bin configuration the ideal one as it contains most information.
However, as we show in appendix \ref{sec:diff_stat} for 12T, 12E bins the constraints on only non-zero $\eps_{\mu\mu}^V$  flatten out for large NSI values which leads to $\eps_{\mu\mu}^V\sim\mathcal{O}(0.8)$ to be allowed at the 70$\%$ C.L. These results are not physical and come from the fact that for large NSI values the pull term for the signal normalization is close to -1 which drastically reduces the number of signal events however only increases the value of the TS by maximally $(-1/0.28)^2$ which is a small change compared to the contribution to the TS from the individual bins which sum up to around 150. One possibility to restore the physicality of the results is to introduce asymmetric pull terms which are steeper for negative values of the normalization. In appendix \ref{sec:diff_stat} we present an alternative parametrization for the pull terms and derive the resultant constraints. 

\begin{figure}
 \centering
 \includegraphics[width=0.6\linewidth]{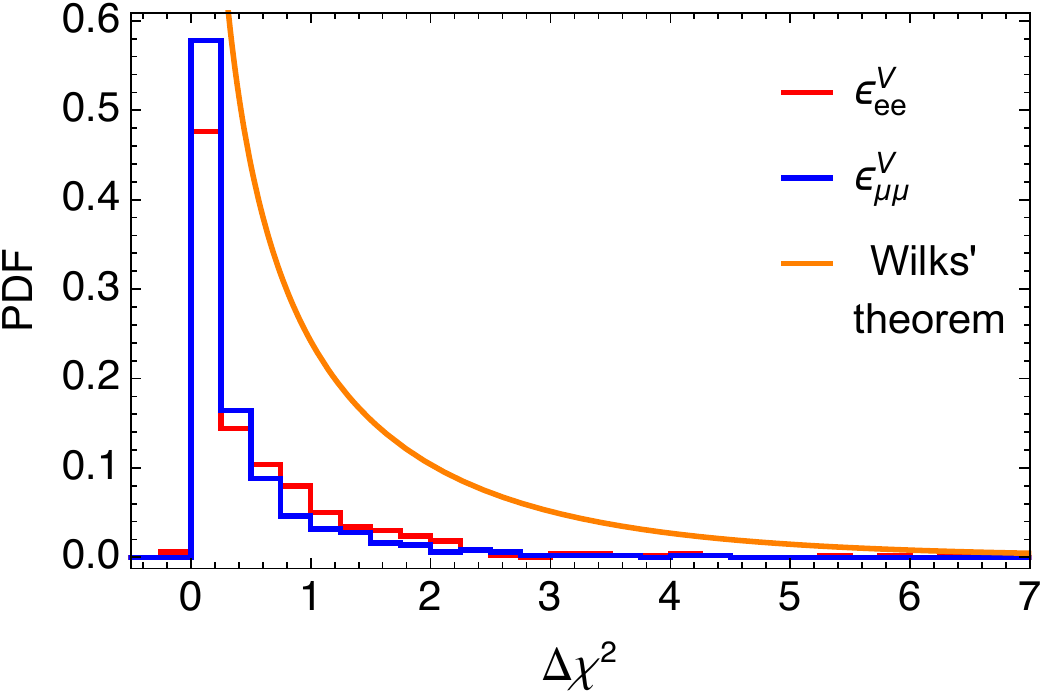}
 \caption{PDF for the best fit values of $\eps_{ee}^V,~\eps_{\mu\mu}^V$, assuming all other NSI parameters to be zero, using 144 bins compared to a $\chi^2$ distribution with 1 dof which is expected to hold if the PDF can be described by Wilks' theorem. }
 \label{fig:pvalwilks}
\end{figure}

Fig.~\ref{fig:pvalwilks} shows another important feature of the CE$\nu$NS data, namely the limitations of Wilks' theorem. 
A common source of error in statistical analysis is to assume that the test statistic follows a $\chi^2$ distribution with mean equal to the degrees of freedom (dof's) according to Wilks' theorem \cite{Wilks:1938dza}. This is often not correct and can lead to significantly incorrect results \cite{Feldman:1997qc,Lyons:2014kta, Algeri:2019arh}, as it has been pointed out recently in the context of short baseline neutrino oscillations \cite{Agostini:2019jup, Silaeva:2020yot, Giunti:2020uhv, Berryman:2020agd,Almazan:2020drb,Qian:2014nha,Coloma:2020ajw} and for the determinations of the leptonic CP phase \cite{Blennow:2014sja,Blennow:2013oma,Schwetz:2006md,Elevant:2015ska}.
In fact for low number of events per bin the test statistic is not expected to follow a $\chi^2$ distribution as in this case the necessary condition to fulfill Wilks' theorem that enough data is observed is not met (see for example \cite{Algeri:2019arh}). 
In order to overcome this problem it is necessary to analyze the CsI data by performing a MC estimation of the distribution of the test statistic defined in eq.~\ref{eq:ts}. While the validity of Wilks' theorem has been checked for the significance of the observation in the case of the most recent COHERENT Argon data \cite{Akimov:2020pdx,Daughhetee2020} using the total number of events  such checks are always important as
the number of detected CE$\nu$NS events is expected to be small in many CE$\nu$NS experiments (in particular those using accelerators as neutrino source). Hence a consistent and correct analysis needs to rely on the use of a MC.

In the following we will hence use the this alternative, asymmetric pull term parametrization.
We show the resultant constraints on the NSI parameters in fig.~\ref{fig:constraints}. 
The best fit points for the diagonal NSI parameters are 
$\eps_{ee}^V= 3.2\cdot 10^{-2},\eps_{\mu\mu}^V= 4.2\cdot 10^{-3}$. The degenerate point with the SM is at $\eps_{\alpha\alpha}^V=-\tfrac{2(g_p^V+g_n^V Y_n)}{3(1+Y_n)}=0.19$ with the neutron fraction of CsI
$Y_n=N_n/N_p=N_n/N_e\approx1.41$. We observe two disjoined allowed regions for $\eps_{\mu\mu}^V$. 
The constraints on $\eps_{e\mu}^V$ and $\eps_{e\tau}^V$ are similar whereas $\eps_{\mu\tau}^V$ is less constrained. The best fit points are
$\eps_{e\mu}^V=7.3\cdot 10^{-5},~\eps_{e\tau}^V= -7.2\cdot 10^{-6},~\eps_{\mu\tau}^V=-3.0\cdot 10^{-5}$.
 While the allowed region for the flavor diagonal NSI parameters is asymmetric and the CsI data prefers larger positive NSI parameters
the constraint on flavor changing NSI parameters is symmetric around $\epsilon_{\alpha\beta}^V\approx 0$ as there is no flavor changing SM contribution to the cross section (see eq.~\eqref{eq:QWnsi}).
The direction of the asymmetry of the flavor diagonal elements is easily understood as $g_n^V\approx-\frac12$.

\subsection{Multiple NSI parameters}
Next we consider the case that all 5 NSI parameters are non-zero simultaneously. In principle  a full 5D Feldman-Cousins approach should be used. However this is extremely computationally expensive. To circumvent this problem we treat the NSI parameters that are not of interest as nuisance parameters and marginalize over them, an approach proposed in \cite{NOVA-doc-15884-v3}. However unlike the normalization nuisance parameters we do not include pull terms for the NSI parameters as we don't have a bias on their real values. Following the approach described in the previous section we obtain the confidence intervals for the parameters of interest.

\begin{figure}
\centering
\includegraphics[width=4in]{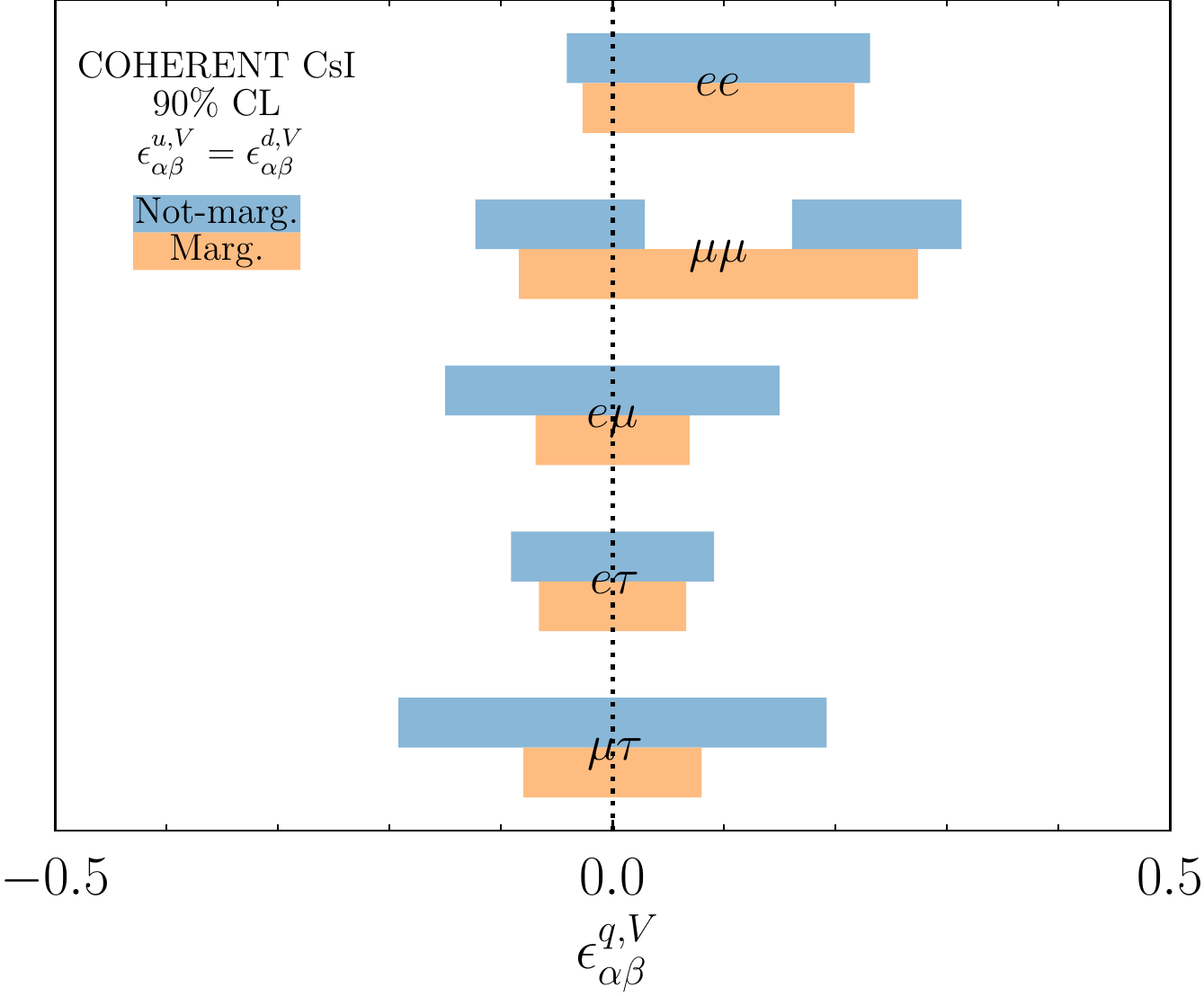}
\caption{The allowed regions at 90\% C.L.
The blue regions represent the one-at-a-time constraints while the orange regions are marginalized over the other NSI parameters.}
\label{fig:constraints}
\end{figure}

In fig.~\ref{fig:constraints} we present the results for the flavor diagonal ($\eps_{ee}^V,~\eps_{\mu\mu}^V$) and flavor off-diagonal NSI parameters ($\eps_{e\mu}^V,~\eps_{e\tau}^V,~\eps_{\mu\tau}^V$) marginalized over the remaining four NSI parameters using 12T, 12E bins, and  asymmetric pull terms.
The best fit value in this case is 
\begin{align}
 \eps_{ee}^V= 3.3\cdot 10^{-2},~
\eps_{\mu\mu}^V= 9.8\cdot 10^{-4},~\nonumber\\
\eps_{\e\mu}^V= - 7.9 \cdot 10^{-3},~
\eps_{\e\tau}^V=9.1\cdot 10^{-5},~
\eps_{\mu\tau}^V= -9.4\cdot 10^{-3} 
\end{align}

Tab.~\ref{tab:nsipval} summarizes our results at 90$\%$ C.L.~and 95$\%$ C.L.~ranges of the NSI parameters derived in this section and in the previous, see also fig.~\ref{fig:constraints}. These constraints can be easily translated to 
constraints on NSI parameters affecting up and down quarks differently by noting that what is measured depends on $(2+Y_n)\eps_{\alpha\beta}^{u,V}+(1+2Y_n)\eps_{\alpha\beta}^{d,V}$
with the neutron fraction $Y_n=N_n/N_p=N_n/N_e\approx1.41$ for CsI \cite{Denton:2018xmq}.
For example, to translate the constraints listed in table \ref{tab:nsipval} to up quark only NSI ($\eps^d=0$) one rescales those constraints by $\frac{2+Y_n}{3(1+Y_n)}=0.47$.
To translate them to down quark only one multiplies them by $\frac{1+2Y_n}{3(1+Y_n)}=0.53$.

We note that the marginalized constraints are often tighter than the one at a time, not marginalized constraints.
This is particularly true for the off-diagonal NSI parameters while for the diagonal ones the difference is less apparent. This indicates that the off-diagonal NSI parameters don't lead to an improvement of the fit.

It should be noted that this result for the off-diagonal NSI parameters using the FC procedure is opposite when assuming Wilks' theorem. In this case the  not marginalized constraints  are tighter than the marginalized ones (see also  ref.~\cite{Giunti:2019xpr} and 
appendix \ref{sec:comparison}). However there is no big difference for the diagonal NSI parameters.

While these results may seem counterintuitive, there are a few reasons why this holds. First, for multiple NSI parameters, the best fit scenario is a better fit to the data. This is
in particular true for the off-diagonal parameters. This means that a comparison between the constraints between the marginalized and not marginalized cases cannot be done on equal footing and can hence be misleading.  As a difference to Wilks' theorem is that the impact of the  goodness of fit plays an important role in a Feldman-Cousins style analysis, this can change the outcome in unintuitive ways. 
It should be also noted that the signal also depends non-trivially on the NSI parameters as $Q_w^2$ is a non-linear function of them. Furthermore, there are also non-trivial degeneracies of the NSI parameters present. Apart from the degeneracies with the SM point in the one parameter scenario already mentioned there are also degeneracies assuming multiple NSI parameters. In fact, 
the effect of the off-diagonal parameters can be compensated by non-zero diagonal parameters until $\epsilon_{\alpha\beta}^V\sim 0.1$. This means 
there is to a change of degrees of freedom around $\epsilon_{\alpha\beta}^V\sim 0.1$.

Another reason for this counter intuitive results could be the effect of nuisance parameter undercoverage in the Feldman-Cousins procedure. As laid out in \cite{10.2307/2290928} and mentioned in sec.~\ref{sec:nsi_one_at_time} nuisance parameter coverage can be tested by varying the values of the nuisance parameters when simulating the experiments to find regions which are not included in the best fit region. Doing this for all nuisance parameters at once is an extraordinary computational task, in particular for the case of the marginalized constraints. As a simplified test we conducted the procedure from \cite{10.2307/2290928}  varying one nuisance parameter at a time for the not marginalized and marginalized NSI constraints. The results indicate that undercoverage could be a reason for the counter intuitive results as we find in particular for the off-diagonal NSI parameters in the marginalized case regions of parameter space not included in the best fit region. However it should be noted that the TS in this case is significantly worse such that these regions are not as a good fit to the data as the best fit regions. As far as we know there is no procedure in the literature to cope with this problem, in fact coverage due to nuisance parameters is rarely tested due to the extreme computational cost. As an example the NOvA experiment tested coverage for the rejection power of the neutrino mass ordering and found no undercoverage \cite{Pershey:2018gtf} however the uncertainties on the nuisance parameters (the oscillation parameters in this case) are smaller than the uncertainties we consider here (in particular the signal normalization uncertainty of $28\%$), indicating that in general for small uncertainties undercoverage is usually not expected.

All of these points help to understand our results and should be kept in mind when interpreting the results and comparing them to other results in the literature.

Our $\eps_{ee}^V$ results are compared to others in the literature in appendix \ref{sec:comparison}. 
Comparing the values allowing only one NSI parameter at a time to other constraints on NSI from oscillations measurements (see \cite{Esteban:2018ppq} for a global analysis) we find that our COHERENT constraints on $\eps_{e\tau}^V$ improve over them while atmospheric oscillation data and long baseline data lead to stronger constraints on $\eps_{e\mu}^V,~\eps_{\mu\tau}^V$ compared to COHERENT CsI.

\begin{table}
\def\tspace{0.07in}
\centering
\caption{ Allowed ranges of the NSI parameters derived using 12T, 12 E bins with a MC simulation and the FC procedure and asymmetric pull terms using the COHERENT CsI data.
The constraints apply to mediator masses $\gtrsim100$ MeV. There are two disjoint regions when allowing only for non-zero $\eps_{\mu\mu}^V$.
These constraints, which are for $\eps^{u,V}=\eps^{d,V}$, can be translated into quark flavor specific NSI, see the text.}
\begin{tabular}{c|c|c|c}
&NSI&$90\%$ C.L.& $95\%$ C.L.\\\hline
\multirow{5}{*}{not marginalized}&$\eps_{ee}^V$&[-0.041, 0.231]&[-0.063, 0.253]\\[\tspace]
&$\eps_{\mu\mu}^V$&[-0.123, 0.029] $\oplus$ [0.161, 0.313]
&[-0.197, 0.037] $\oplus$ [0.153, 0.387]
\\[\tspace]
&$\eps_{e\mu}^V$&[-0.150, 0.150]&[-0.298, 0.298]\\[\tspace]
&$\eps_{e\tau}^V$&[-0.091, 0.091]&[-0.116, 0.116]\\[\tspace]
&$\eps_{\mu\tau}^V$&[-0.192, 0.192]&[-0.319, 0.319]\\[\tspace]
\hline
\multirow{5}{*}{marginalized}&$\eps_{ee}^V$&[-0.027, 0.217]&[-0.061, 0.251]\\[\tspace]
&$\eps_{\mu\mu}^V$&[-0.084, 0.274] 
&[-0.246, 0.436]
\\[\tspace]
&$\eps_{e\mu}^V$&[-0.069, 0.069]&[-0.090, 0.090]\\[\tspace]
&$\eps_{e\tau}^V$&[-0.066, 0.066]&[-0.089, 0.089]\\[\tspace]
&$\eps_{\mu\tau}^V$&[-0.080, 0.080]&[-0.089, 0.089]\\[\tspace]
\end{tabular}
\label{tab:nsipval}
\end{table}

\subsection{The impact of more statistics}
As we have demonstrated in the previous sections Wilks' theorem is not applicable for the current CsI data. However one might wonder if with more statistics the requirements for Wilks' theorem are fulfilled. Furthermore, it is instructive to see if for future data less bins still lead to worse constraints than 144 bins.

\begin{figure}
 \centering
 \includegraphics[width=0.6\linewidth]{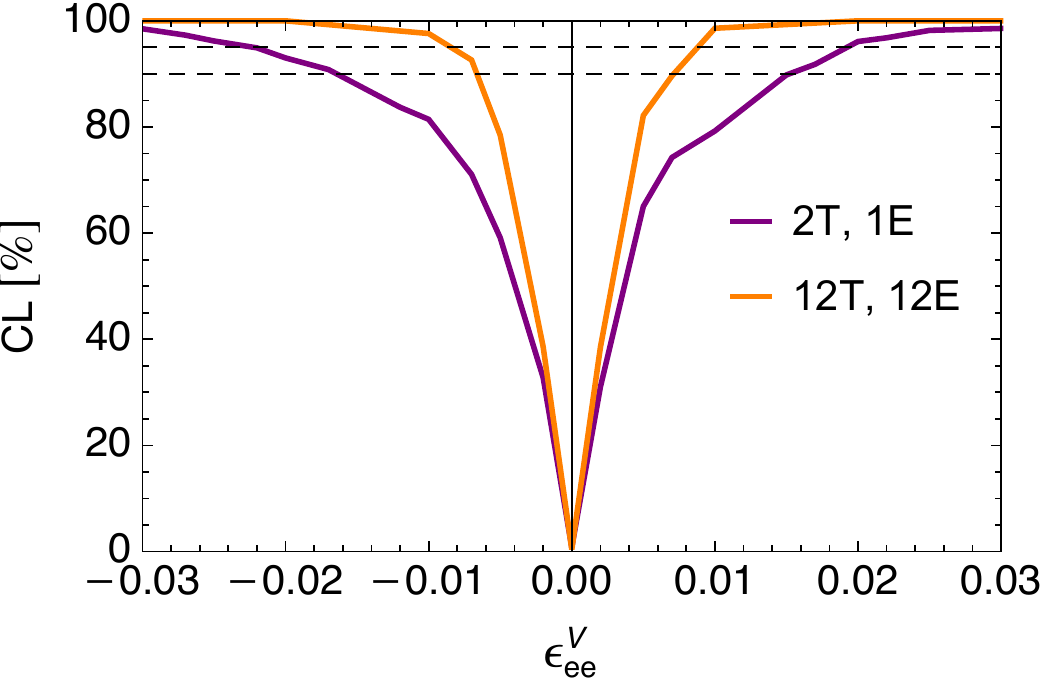}
 \caption{Forecasted sensitivity assuming as real data the SM with 100 times more exposure using the current experimental configuration and 250 background events in each of the 144 bins. The confidence regions have been calculated using 2T, 1E bins (purple) or 144 bins (orange) using a MC simulation. }
 \label{fig:pvalnsifuture}
\end{figure}

To answer this questions we simulate 100 times more CsI data with the current experimental configuration, assuming the SM will be measured. For the background we conservatively assume a flat distribution of 250 events per bin in each of the 144 bins. We calculate the TS assuming only $\eps_{ee}^V $ to be non-zero. In fig.~\ref{fig:pvalnsifuture} we show the results using two timing and one energy bin,  and 144 bins. 
Even with more statistics the conclusions for the ideal analysis strategy derived from the current data are still valid, namely that 12T, 12E bins provide the most information. Furthermore, we have checked that also in the case of more statistics Wilks' theorem is not fulfilled.
These results further demonstrate that independent of the shape of the background data Wilks' theorem fails. The reasons for this might be the non-linear dependence of the signal on the NSI parameters, $Q_w$ depends non-linearly on them but also smearing might effect the validity of Wilks' theorem.

\section{Summary and Conclusion}
\label{sec:conc}
The first observation of CE$\nu$NS in 2017 has sparked large interest 
to use this process to constrain Standard Model parameters as well as to test a broad range of new physics scenarios. However to obtain statistically meaningful constraints from CE$\nu$NS a correct analysis procedure needs to be developed. To this end in this manuscript we 
revisited the analysis of CE$\nu$NS and established an adequate analysis procedure to obtain viable results. 
As concrete examples we show the difference between our statistical approach and commonly used incorrect approaches for the case of the SM prediction as well as for non standard interactions.
We focused on the publicly available data from the COHERENT CsI observation of CE$\nu$NS in 2017 as an example, however our results apply to a wide class of CE$\nu$NS experiments.

First, we identified two general subtleties of the CE$\nu$NS analysis
which are shortcomings of previous works,
namely the effect of smearing and the choice of number of bins.
Since many CE$\nu$NS detectors measure the number of PE instead of nuclear recoil directly we first pointed out that smearing of the predicted events needs to be taken into account. This accounts for the 
probability that sometimes a nuclear recoil event yields a different number of PE than average. Ideally smearing of the number of events via a Poisson distribution which translates the number of events in raw PE bins to the number of events in true, detected PE bins
is applied to the sum of signal and background events. 
However we found that on the publicly released CsI background data cuts have been already imposed as we could not find a good fit of an event distribution in terms of raw PE to the provided distribution in true PE.
Furthermore, unlike in other analyses, we use the sum of all three background data files to increase the statistics of the background.

We then investigated the question of ideal number of bins. As background fluctuations can mimic the signal choosing large bins to increase the number of events per bin and hence decrease the Poisson fluctuations per bin is desirable however if new physics predicts a certain timing or energy behavior too large bins can decrease the sensitivity of the analysis. 
Therefore, the ideal number (and width) of bins depend on the model one wants to test. As we focused in this manuscript on the case of flavor specific NSI with a heavy mediator (with mass above 100 MeV) we find that using 12 timing and 12 energy bins contain most information and leads to the best constraints. 
On the other hand, we find that for all bin configurations the SM is a good fit to the data.

Additionally, we emphasized for the first time in the literature that Wilks' theorem is not fulfilled for BSM analyses of the COHERENT CE$\nu$NS data. This statement is even true for a increased exposure by a factor of 100. Hence a MC estimation of the confidence levels is necessary. Using this improved statistical framework we presented the results for NSI parameters allowing only one of them to be non-zero at the time or considering all of them to be non-zero simultaneously. 
In all cases our results provide the most statistically sound constraints in the literature emphasizing the necessity of the improved analysis framework.

Our results show that using the correct statistical approach as well as accounting for experimental subtlety like smearing, the use of all available background data and the choice of binning has an important impact on the results, both for the SM but also new physics models.

Our proposed analysis approach is accompanied by the first publicly released code to calculate and analyze CE$\nu$NS. The code allows for an easy use of our analysis as smearing, the background template and the calculation of the $p$ value and the FC approach are already implemented and can be extended in a straight forward way to probe other new physics scenarios.
Furthermore, we provide our numerical results in the form of datafiles.

\acknowledgments
We thank Carlos Arg\"uelles, Alexey Konovalov, Daniel Pershey, and Grayson Rich for helpful conversations.
We also thank our anonymous referee.
This work is supported by the US Department of Energy under Grant Contract DE-SC0012704.

\appendix

\section{The Impact of Different Pull Terms}
\label{sec:diff_stat}
\begin{figure}
\centering
\includegraphics[width=0.47\linewidth]{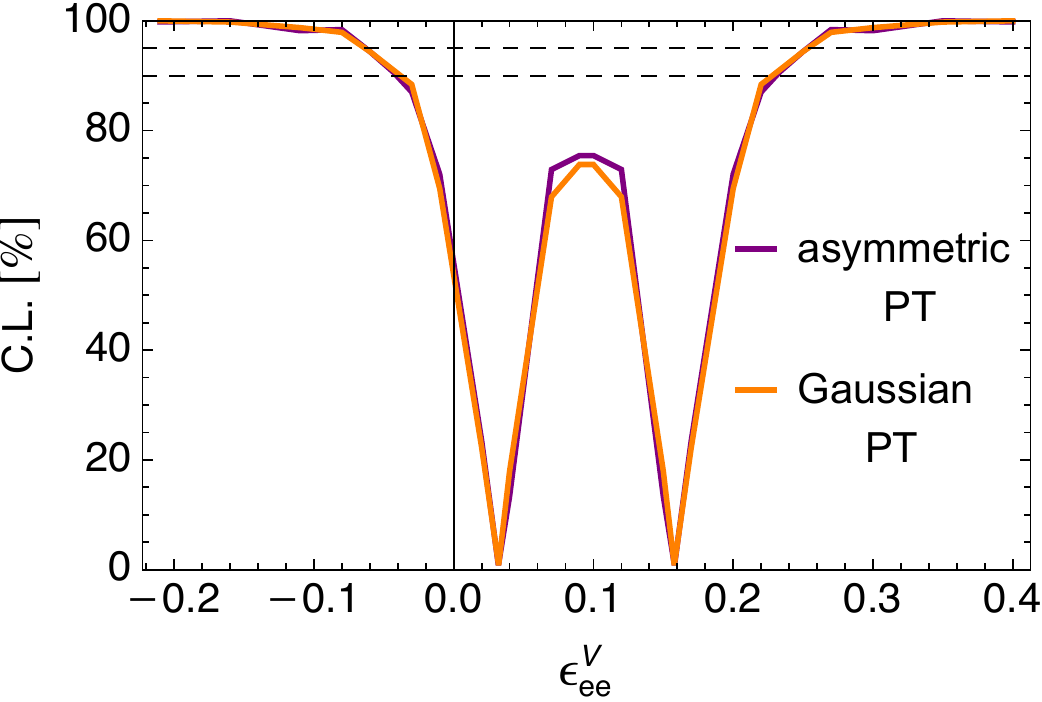}\hspace{0.5cm}
\includegraphics[width=0.47\linewidth]{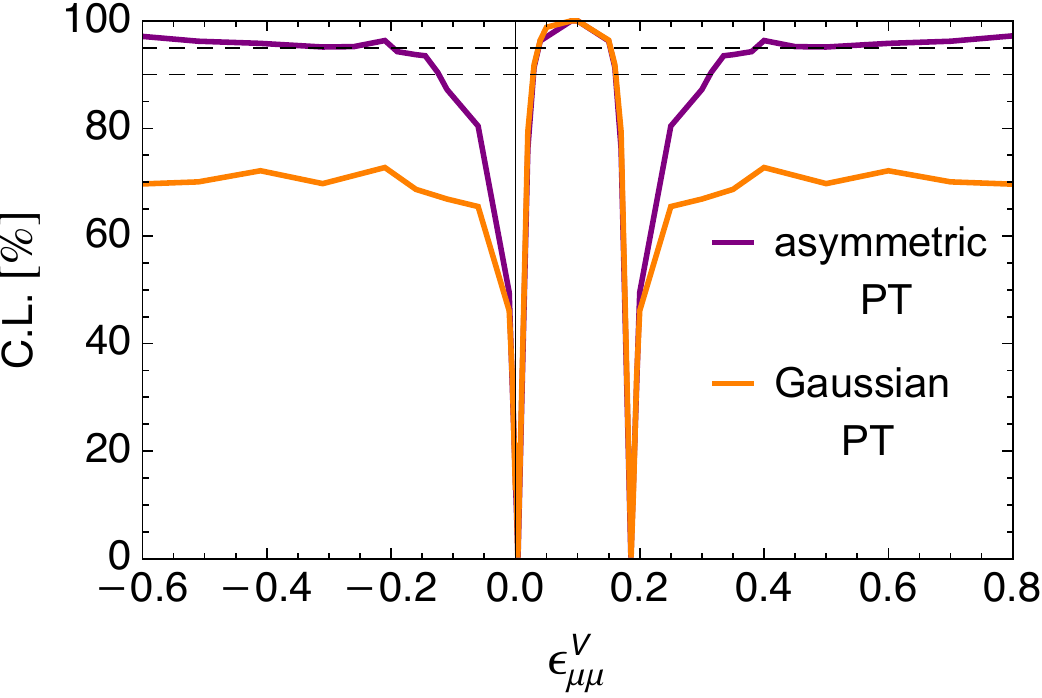}
\caption{Constraint on $\eps_{ee}^V,~\eps_{\mu\mu}^V$ assuming the remaining NSI parameter to be zero using Gaussian pull terms (orange lines) or the new parametrization for the pull terms (purple) using 144 bins. }
\label{fig:pulls}
\end{figure}

As shown in fig.~\ref{fig:pvalnsi} an analysis using 144 bins contains most information compared to an analysis using fewer bins. However, as we show in  fig.~\ref{fig:pulls} the constraints derived when 144 bins flatten out for larger values of $\eps_{\mu\mu}^V$ parameters, which is not the case for $\eps_{ee}^V$. This can be easily understood from the fact that $\eps_{\mu\mu}^V$ affects all timing bins, including the prompt timing bins which have the most statistic and hence drive the TS. 
The flattening of the constraints is due to the signal normalization pull terms which get very small to suppress the signal however the according pull terms do not lead to a large enough penalty in this case. For large $\eps_{\mu\mu}^V$ a good fit can only be achieved if the signal in the prompt and delayed bins is very suppressed which is not the case for $\eps_{ee}^V$ which only affects the delayed bins. The flattening has not been noticed before by other authors who assumed Wilks' theorem to hold as this effect appears around $\Delta \chi^2\approx 13$ which corresponds to a significance $>3\sigma$. 

One way to reconstitute physical results is to introduce asymmetric pull terms instead of assuming them to be Gaussian. 
As already discussed in section \ref{sec:analysis} for a small number of signal events non-Gaussian pull terms are expected.
In order to maintain the Gaussian behavior for small values of the normalizations but introduce a larger penalty for large negative values we use as new pull term parametrization
\begin{align}
 f_{\text{new~pull}}(x,\sigma_x)=\frac{2}{\sigma_x^2}(x - \log(x + 1))\,.
\end{align}
This pull term is derived from a rescaled Poisson.

As expected with the new pull term parametrization the flattening of the constraints for large NSI parameters is avoided, leading to physically sensible results. 
We should stress that while the new pull term results in more ``sensible'' results, there is no guarantee that either is correct.
In practice, one should obtain the probability density function for the constraint used in a given systematic and to be cautious when a pull is going outside the region provided.

\section{Comparison to other analyses in the literature}
\label{sec:comparison}
\begin{figure}
\centering
\includegraphics[width=4in]{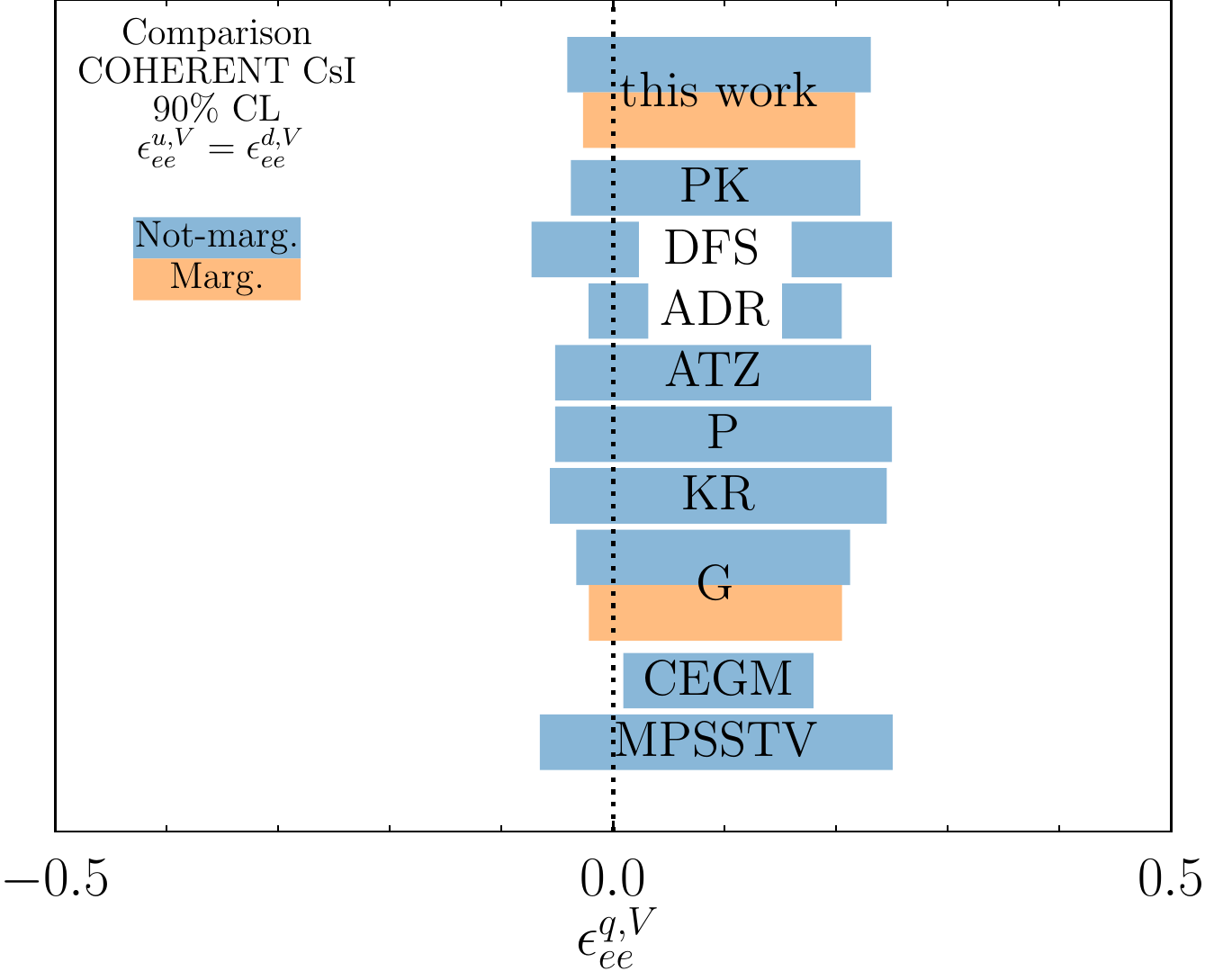}
\caption{The allowed regions for $\epsilon_{ee}^V$ at 90\% C.L.~assuming $\epsilon_{ee}^{u,V}=\epsilon_{ee}^{d,V}$ from this work and PK \cite{Kosmas:2017tsq}, DFS \cite{Denton:2018xmq}, ADR \cite{AristizabalSierra:2018eqm}, ATZ \cite{Altmannshofer:2018xyo}, P \cite{Papoulias:2019txv}, KR \cite{Khan:2019cvi}, G \cite{Giunti:2019xpr}, CEGM \cite{Coloma:2019mbs}, and MPSSTV \cite{Miranda:2020tif}.}
\label{fig:comparison}
\end{figure}

Several other analyses of COHERENT's CsI data in the context of NSIs have been performed in the literature and here we compare our results to several others\footnote{Note that several other analyses do not present specifically COHERENT constraints or do so only for certain combinations of flavor at a time and are not shown here \cite{Akimov:2017ade,Coloma:2017ncl,Liao:2017uzy,Dent:2017mpr,Billard:2018jnl,Esteban:2018ppq,Heeck:2018nzc}.}.
We stress that a tighter constraint is not a \emph{better} constraint.
Our goal is to highlight the size of the effect that different analyses have on the overall constraint.
In fig.~\ref{fig:comparison} we show the comparison of the $ee$ NSI term after making the necessary translations from other papers; taking $\Delta\chi^2=2.71$ $\Rightarrow$ 90\% C.L.~and converting quark specific couplings to our scenario of $\eps^u=\eps^d$ as necessary\footnote{For a single material this second translation provides no issues. When combining data from multiple NSI constraints, a translation from one assumption about the relative couplings to different quarks to another assumption cannot be done without performing a re-analysis.}.
The different papers shown are listed here:
\begin{itemize}
\item Ref.~\cite{Kosmas:2017tsq} by Papoulias and Kosmas (PK) examined several BSM scenarios in the context of COHERENT.
\item Ref.~\cite{Denton:2018xmq} by Denton, Farzan, and Shoemaker (DFS) examined the light mediator case in the context of the LMA-Dark degenerate point.
\item Ref.~\cite{AristizabalSierra:2018eqm} by Aristizabal, De Romeri, and Rojas (ADR) examined different Lorentz structures of new interactions.
\item Ref.~\cite{Altmannshofer:2018xyo} by Altmannshofer, Tammaro, and Zupan (ATZ) examined NSIs in EFT and SMEFT frameworks from multiple low energy experiments.
\item Ref.~\cite{Papoulias:2019txv} by Papoulias (P) examined the impact of a new quenching factor measurement.
\item Ref.~\cite{Khan:2019cvi} by Kahn and Rodejohann (KR) also examined the impact of a new quenching factor measurement.
\item Ref.~\cite{Giunti:2019xpr} by Giunti (G) examined NSI constraints from COHERENT in the heavy mediator case.
\item Ref.~\cite{Coloma:2019mbs} by Coloma, Esteban, Gonzalez-Garcia, and Maltoni (CEGM) combined COHERENT constraints on NSIs along with that from oscillations.
\item Ref.~\cite{Miranda:2020tif} by Miranda, Papoulias, Sanchez Garcia, Sanders, T\'ortola, and Valle estimated the sensitivity to NSIs with LAr data at COHERENT.
\end{itemize}

Beyond differences in statistical treatment with regards to Wilks' theorem, these analyses differ in numerous ways including the number of bins used, the treatment of backgrounds, the performance of smearing, the form factors used, and the quenching factor details.

\bibliographystyle{JHEP}
\bibliography{bibfile}

\providecommand{\href}[2]{#2}\begingroup\raggedright\begin{thebibliography}{100}

\bibitem{Freedman:1973yd}
D.~Z. Freedman, \emph{{Coherent Neutrino Nucleus Scattering as a Probe of the
  Weak Neutral Current}},
  \href{http://dx.doi.org/10.1103/PhysRevD.9.1389}{\emph{Phys. Rev.} {\bf D9}
  (1974) 1389--1392}.

\bibitem{Drukier:1983gj}
A.~Drukier and L.~Stodolsky, \emph{{Principles and Applications of a Neutral
  Current Detector for Neutrino Physics and Astronomy}},
  \href{http://dx.doi.org/10.1103/PhysRevD.30.2295}{\emph{Phys. Rev.} {\bf D30}
  (1984) 2295}.

\bibitem{Akimov:2017ade}
{\scshape COHERENT} collaboration, D.~Akimov et~al., \emph{{Observation of
  Coherent Elastic Neutrino-Nucleus Scattering}},
  \href{http://dx.doi.org/10.1126/science.aao0990}{\emph{Science} {\bf 357}
  (2017) 1123--1126}, [\href{https://arxiv.org/abs/1708.01294}{{\tt
  1708.01294}}].

\bibitem{Dent:2016wcr}
J.~B. Dent, B.~Dutta, S.~Liao, J.~L. Newstead, L.~E. Strigari and J.~W. Walker,
  \emph{{Probing light mediators at ultralow threshold energies with coherent
  elastic neutrino-nucleus scattering}},
  \href{http://dx.doi.org/10.1103/PhysRevD.96.095007}{\emph{Phys. Rev. D} {\bf
  96} (2017) 095007}, [\href{https://arxiv.org/abs/1612.06350}{{\tt
  1612.06350}}].

\bibitem{Coloma:2017egw}
P.~Coloma, P.~B. Denton, M.~Gonzalez-Garcia, M.~Maltoni and T.~Schwetz,
  \emph{{Curtailing the Dark Side in Non-Standard Neutrino Interactions}},
  \href{http://dx.doi.org/10.1007/JHEP04(2017)116}{\emph{JHEP} {\bf 04} (2017)
  116}, [\href{https://arxiv.org/abs/1701.04828}{{\tt 1701.04828}}].

\bibitem{Ge:2017mcq}
S.-F. Ge and I.~M. Shoemaker, \emph{{Constraining Photon Portal Dark Matter
  with Texono and Coherent Data}},
  \href{http://dx.doi.org/10.1007/JHEP11(2018)066}{\emph{JHEP} {\bf 11} (2018)
  066}, [\href{https://arxiv.org/abs/1710.10889}{{\tt 1710.10889}}].

\bibitem{Coloma:2017ncl}
P.~Coloma, M.~Gonzalez-Garcia, M.~Maltoni and T.~Schwetz, \emph{{COHERENT
  Enlightenment of the Neutrino Dark Side}},
  \href{http://dx.doi.org/10.1103/PhysRevD.96.115007}{\emph{Phys. Rev. D} {\bf
  96} (2017) 115007}, [\href{https://arxiv.org/abs/1708.02899}{{\tt
  1708.02899}}].

\bibitem{Liao:2017uzy}
J.~Liao and D.~Marfatia, \emph{{COHERENT constraints on nonstandard neutrino
  interactions}},
  \href{http://dx.doi.org/10.1016/j.physletb.2017.10.046}{\emph{Phys. Lett. B}
  {\bf 775} (2017) 54--57}, [\href{https://arxiv.org/abs/1708.04255}{{\tt
  1708.04255}}].

\bibitem{Dent:2017mpr}
J.~B. Dent, B.~Dutta, S.~Liao, J.~L. Newstead, L.~E. Strigari and J.~W. Walker,
  \emph{{Accelerator and reactor complementarity in coherent neutrino-nucleus
  scattering}}, \href{http://dx.doi.org/10.1103/PhysRevD.97.035009}{\emph{Phys.
  Rev. D} {\bf 97} (2018) 035009},
  [\href{https://arxiv.org/abs/1711.03521}{{\tt 1711.03521}}].

\bibitem{Lindner:2016wff}
M.~Lindner, W.~Rodejohann and X.-J. Xu, \emph{{Coherent Neutrino-Nucleus
  Scattering and new Neutrino Interactions}},
  \href{http://dx.doi.org/10.1007/JHEP03(2017)097}{\emph{JHEP} {\bf 03} (2017)
  097}, [\href{https://arxiv.org/abs/1612.04150}{{\tt 1612.04150}}].

\bibitem{Abdullah:2018ykz}
M.~Abdullah, J.~B. Dent, B.~Dutta, G.~L. Kane, S.~Liao and L.~E. Strigari,
  \emph{{Coherent elastic neutrino nucleus scattering as a probe of a Z'
  through kinetic and mass mixing effects}},
  \href{http://dx.doi.org/10.1103/PhysRevD.98.015005}{\emph{Phys. Rev. D} {\bf
  98} (2018) 015005}, [\href{https://arxiv.org/abs/1803.01224}{{\tt
  1803.01224}}].

\bibitem{Shoemaker:2017lzs}
I.~M. Shoemaker, \emph{{COHERENT search strategy for beyond standard model
  neutrino interactions}},
  \href{http://dx.doi.org/10.1103/PhysRevD.95.115028}{\emph{Phys. Rev. D} {\bf
  95} (2017) 115028}, [\href{https://arxiv.org/abs/1703.05774}{{\tt
  1703.05774}}].

\bibitem{Kosmas:2017tsq}
D.~Papoulias and T.~Kosmas, \emph{{COHERENT constraints to conventional and
  exotic neutrino physics}},
  \href{http://dx.doi.org/10.1103/PhysRevD.97.033003}{\emph{Phys. Rev. D} {\bf
  97} (2018) 033003}, [\href{https://arxiv.org/abs/1711.09773}{{\tt
  1711.09773}}].

\bibitem{Farzan:2018gtr}
Y.~Farzan, M.~Lindner, W.~Rodejohann and X.-J. Xu, \emph{{Probing neutrino
  coupling to a light scalar with coherent neutrino scattering}},
  \href{http://dx.doi.org/10.1007/JHEP05(2018)066}{\emph{JHEP} {\bf 05} (2018)
  066}, [\href{https://arxiv.org/abs/1802.05171}{{\tt 1802.05171}}].

\bibitem{Brdar:2018qqj}
V.~Brdar, W.~Rodejohann and X.-J. Xu, \emph{{Producing a new Fermion in
  Coherent Elastic Neutrino-Nucleus Scattering: from Neutrino Mass to Dark
  Matter}}, \href{http://dx.doi.org/10.1007/JHEP12(2018)024}{\emph{JHEP} {\bf
  12} (2018) 024}, [\href{https://arxiv.org/abs/1810.03626}{{\tt 1810.03626}}].

\bibitem{Datta:2018xty}
A.~Datta, B.~Dutta, S.~Liao, D.~Marfatia and L.~E. Strigari, \emph{{Neutrino
  scattering and B anomalies from hidden sector portals}},
  \href{http://dx.doi.org/10.1007/JHEP01(2019)091}{\emph{JHEP} {\bf 01} (2019)
  091}, [\href{https://arxiv.org/abs/1808.02611}{{\tt 1808.02611}}].

\bibitem{Kosmas:2017zbh}
T.~Kosmas, D.~Papoulias, M.~Tortola and J.~Valle, \emph{{Probing light sterile
  neutrino signatures at reactor and Spallation Neutron Source neutrino
  experiments}},
  \href{http://dx.doi.org/10.1103/PhysRevD.96.063013}{\emph{Phys. Rev. D} {\bf
  96} (2017) 063013}, [\href{https://arxiv.org/abs/1703.00054}{{\tt
  1703.00054}}].

\bibitem{Blanco:2019vyp}
C.~Blanco, D.~Hooper and P.~Machado, \emph{{Constraining Sterile Neutrino
  Interpretations of the LSND and MiniBooNE Anomalies with Coherent Neutrino
  Scattering Experiments}},
  \href{http://dx.doi.org/10.1103/PhysRevD.101.075051}{\emph{Phys. Rev. D} {\bf
  101} (2020) 075051}, [\href{https://arxiv.org/abs/1901.08094}{{\tt
  1901.08094}}].

\bibitem{Denton:2018xmq}
P.~B. Denton, Y.~Farzan and I.~M. Shoemaker, \emph{{Testing large non-standard
  neutrino interactions with arbitrary mediator mass after COHERENT data}},
  \href{http://dx.doi.org/10.1007/JHEP07(2018)037}{\emph{JHEP} {\bf 07} (2018)
  037}, [\href{https://arxiv.org/abs/1804.03660}{{\tt 1804.03660}}].

\bibitem{Canas:2019fjw}
B.~Canas, E.~Garces, O.~Miranda, A.~Parada and G.~Sanchez~Garcia,
  \emph{{Interplay between nonstandard and nuclear constraints in coherent
  elastic neutrino-nucleus scattering experiments}},
  \href{http://dx.doi.org/10.1103/PhysRevD.101.035012}{\emph{Phys. Rev. D} {\bf
  101} (2020) 035012}, [\href{https://arxiv.org/abs/1911.09831}{{\tt
  1911.09831}}].

\bibitem{Chang:2020jwl}
W.-F. Chang and J.~Liao, \emph{{Constraints on light singlet fermion
  interactions from coherent elastic neutrino-nucleus scattering}},
  \href{https://arxiv.org/abs/2002.10275}{{\tt 2002.10275}}.

\bibitem{Flores:2020lji}
L.~Flores, N.~Nath and E.~Peinado, \emph{{Non-standard neutrino interactions in
  $U(1)'$ model after COHERENT data}},
  \href{https://arxiv.org/abs/2002.12342}{{\tt 2002.12342}}.

\bibitem{Abdullah:2020iiv}
M.~Abdullah, D.~Aristizabal~Sierra, B.~Dutta and L.~E. Strigari,
  \emph{{Coherent Elastic Neutrino-Nucleus Scattering with directional
  detectors}},  \href{https://arxiv.org/abs/2003.11510}{{\tt 2003.11510}}.

\bibitem{Miranda:2020tif}
O.~Miranda, D.~Papoulias, G.~S. Garcia, O.~Sanders, M.~Tórtola and J.~Valle,
  \emph{{Implications of the first detection of coherent elastic
  neutrino-nucleus scattering (CEvNS) with Liquid Argon}},
  \href{https://arxiv.org/abs/2003.12050}{{\tt 2003.12050}}.

\bibitem{Li:2020lba}
T.~Li, X.-D. Ma and M.~A. Schmidt, \emph{{Generic neutrino interactions with
  sterile neutrinos in light of neutrino-nucleus coherent scattering and meson
  invisible decays}},  \href{https://arxiv.org/abs/2005.01543}{{\tt
  2005.01543}}.

\bibitem{Bowen:2020unj}
M.~Bowen and P.~Huber, \emph{{Reactor neutrino applications and coherent
  elastic neutrino nucleus scattering}},
  \href{https://arxiv.org/abs/2005.10907}{{\tt 2005.10907}}.

\bibitem{Hurtado:2020vlj}
N.~Hurtado, H.~Mir, I.~M. Shoemaker, E.~Welch and J.~Wyenberg, \emph{{Dark
  Matter-Neutrino Interconversion at COHERENT, Direct Detection, and the Early
  Universe}},  \href{https://arxiv.org/abs/2005.13384}{{\tt 2005.13384}}.

\bibitem{Billard:2018jnl}
J.~Billard, J.~Johnston and B.~J. Kavanagh, \emph{{Prospects for exploring New
  Physics in Coherent Elastic Neutrino-Nucleus Scattering}},
  \href{http://dx.doi.org/10.1088/1475-7516/2018/11/016}{\emph{JCAP} {\bf 11}
  (2018) 016}, [\href{https://arxiv.org/abs/1805.01798}{{\tt 1805.01798}}].

\bibitem{AristizabalSierra:2018eqm}
D.~Aristizabal~Sierra, V.~De~Romeri and N.~Rojas, \emph{{COHERENT analysis of
  neutrino generalized interactions}},
  \href{http://dx.doi.org/10.1103/PhysRevD.98.075018}{\emph{Phys. Rev. D} {\bf
  98} (2018) 075018}, [\href{https://arxiv.org/abs/1806.07424}{{\tt
  1806.07424}}].

\bibitem{Bednyakov:2018mjd}
V.~A. Bednyakov and D.~V. Naumov, \emph{{Coherency and incoherency in
  neutrino-nucleus elastic and inelastic scattering}},
  \href{http://dx.doi.org/10.1103/PhysRevD.98.053004}{\emph{Phys. Rev. D} {\bf
  98} (2018) 053004}, [\href{https://arxiv.org/abs/1806.08768}{{\tt
  1806.08768}}].

\bibitem{Akhmedov:2018wlf}
E.~Akhmedov, G.~Arcadi, M.~Lindner and S.~Vogl, \emph{{Coherent scattering and
  macroscopic coherence: Implications for neutrino, dark matter and axion
  detection}}, \href{http://dx.doi.org/10.1007/JHEP10(2018)045}{\emph{JHEP}
  {\bf 10} (2018) 045}, [\href{https://arxiv.org/abs/1806.10962}{{\tt
  1806.10962}}].

\bibitem{Cadeddu:2018dux}
M.~Cadeddu, C.~Giunti, K.~Kouzakov, Y.~Li, A.~Studenikin and Y.~Zhang,
  \emph{{Neutrino Charge Radii from COHERENT Elastic Neutrino-Nucleus
  Scattering}}, \href{http://dx.doi.org/10.1103/PhysRevD.98.113010}{\emph{Phys.
  Rev. D} {\bf 98} (2018) 113010},
  [\href{https://arxiv.org/abs/1810.05606}{{\tt 1810.05606}}].

\bibitem{Heeck:2018nzc}
J.~Heeck, M.~Lindner, W.~Rodejohann and S.~Vogl, \emph{{Non-Standard Neutrino
  Interactions and Neutral Gauge Bosons}},
  \href{http://dx.doi.org/10.21468/SciPostPhys.6.3.038}{\emph{SciPost Phys.}
  {\bf 6} (2019) 038}, [\href{https://arxiv.org/abs/1812.04067}{{\tt
  1812.04067}}].

\bibitem{Altmannshofer:2018xyo}
W.~Altmannshofer, M.~Tammaro and J.~Zupan, \emph{{Non-standard neutrino
  interactions and low energy experiments}},
  \href{http://dx.doi.org/10.1007/JHEP09(2019)083}{\emph{JHEP} {\bf 09} (2019)
  083}, [\href{https://arxiv.org/abs/1812.02778}{{\tt 1812.02778}}].

\bibitem{AristizabalSierra:2019zmy}
D.~Aristizabal~Sierra, J.~Liao and D.~Marfatia, \emph{{Impact of form factor
  uncertainties on interpretations of coherent elastic neutrino-nucleus
  scattering data}},
  \href{http://dx.doi.org/10.1007/JHEP06(2019)141}{\emph{JHEP} {\bf 06} (2019)
  141}, [\href{https://arxiv.org/abs/1902.07398}{{\tt 1902.07398}}].

\bibitem{Cadeddu:2017etk}
M.~Cadeddu, C.~Giunti, Y.~Li and Y.~Zhang, \emph{{Average CsI neutron density
  distribution from COHERENT data}},
  \href{http://dx.doi.org/10.1103/PhysRevLett.120.072501}{\emph{Phys. Rev.
  Lett.} {\bf 120} (2018) 072501},
  [\href{https://arxiv.org/abs/1710.02730}{{\tt 1710.02730}}].

\bibitem{Ciuffoli:2018qem}
E.~Ciuffoli, J.~Evslin, Q.~Fu and J.~Tang, \emph{{Extracting nuclear form
  factors with coherent neutrino scattering}},
  \href{http://dx.doi.org/10.1103/PhysRevD.97.113003}{\emph{Phys. Rev. D} {\bf
  97} (2018) 113003}, [\href{https://arxiv.org/abs/1801.02166}{{\tt
  1801.02166}}].

\bibitem{Huang:2019ene}
X.-R. Huang and L.-W. Chen, \emph{{Neutron Skin in CsI and Low-Energy Effective
  Weak Mixing Angle from COHERENT Data}},
  \href{http://dx.doi.org/10.1103/PhysRevD.100.071301}{\emph{Phys. Rev. D} {\bf
  100} (2019) 071301}, [\href{https://arxiv.org/abs/1902.07625}{{\tt
  1902.07625}}].

\bibitem{Papoulias:2019lfi}
D.~Papoulias, T.~Kosmas, R.~Sahu, V.~Kota and M.~Hota, \emph{{Constraining
  nuclear physics parameters with current and future COHERENT data}},
  \href{http://dx.doi.org/10.1016/j.physletb.2019.135133}{\emph{Phys. Lett. B}
  {\bf 800} (2020) 135133}, [\href{https://arxiv.org/abs/1903.03722}{{\tt
  1903.03722}}].

\bibitem{Dutta:2020che}
B.~Dutta, R.~F. Lang, S.~Liao, S.~Sinha, L.~Strigari and A.~Thompson, \emph{{A
  global analysis strategy to resolve neutrino NSI degeneracies with scattering
  and oscillation data}},  \href{https://arxiv.org/abs/2002.03066}{{\tt
  2002.03066}}.

\bibitem{Dutta:2019eml}
B.~Dutta, S.~Liao, S.~Sinha and L.~E. Strigari, \emph{{Searching for Beyond the
  Standard Model Physics with COHERENT Energy and Timing Data}},
  \href{http://dx.doi.org/10.1103/PhysRevLett.123.061801}{\emph{Phys. Rev.
  Lett.} {\bf 123} (2019) 061801},
  [\href{https://arxiv.org/abs/1903.10666}{{\tt 1903.10666}}].

\bibitem{Bednyakov:2019dbl}
V.~A. Bednyakov and D.~V. Naumov, \emph{{On coherent neutrino and antineutrino
  scattering off nuclei}},
  \href{http://dx.doi.org/10.1134/S1547477119060396}{\emph{Phys. Part. Nucl.
  Lett.} {\bf 16} (2019) 638--646},
  [\href{https://arxiv.org/abs/1904.03119}{{\tt 1904.03119}}].

\bibitem{Miranda:2019wdy}
O.~Miranda, D.~Papoulias, M.~Tórtola and J.~Valle, \emph{{Probing neutrino
  transition magnetic moments with coherent elastic neutrino-nucleus
  scattering}}, \href{http://dx.doi.org/10.1007/JHEP07(2019)103}{\emph{JHEP}
  {\bf 07} (2019) 103}, [\href{https://arxiv.org/abs/1905.03750}{{\tt
  1905.03750}}].

\bibitem{AristizabalSierra:2019ufd}
D.~Aristizabal~Sierra, V.~De~Romeri and N.~Rojas, \emph{{CP violating effects
  in coherent elastic neutrino-nucleus scattering processes}},
  \href{http://dx.doi.org/10.1007/JHEP09(2019)069}{\emph{JHEP} {\bf 09} (2019)
  069}, [\href{https://arxiv.org/abs/1906.01156}{{\tt 1906.01156}}].

\bibitem{Canas:2018rng}
B.~Cañas, E.~Garcés, O.~Miranda and A.~Parada, \emph{{Future perspectives for
  a weak mixing angle measurement in coherent elastic neutrino nucleus
  scattering experiments}},
  \href{http://dx.doi.org/10.1016/j.physletb.2018.07.049}{\emph{Phys. Lett. B}
  {\bf 784} (2018) 159--162}, [\href{https://arxiv.org/abs/1806.01310}{{\tt
  1806.01310}}].

\bibitem{Cadeddu:2018izq}
M.~Cadeddu and F.~Dordei, \emph{{Reinterpreting the weak mixing angle from
  atomic parity violation in view of the Cs neutron rms radius measurement from
  COHERENT}}, \href{http://dx.doi.org/10.1103/PhysRevD.99.033010}{\emph{Phys.
  Rev. D} {\bf 99} (2019) 033010},
  [\href{https://arxiv.org/abs/1808.10202}{{\tt 1808.10202}}].

\bibitem{Dutta:2019nbn}
B.~Dutta, D.~Kim, S.~Liao, J.-C. Park, S.~Shin and L.~E. Strigari, \emph{{Dark
  matter signals from timing spectra at neutrino experiments}},
  \href{http://dx.doi.org/10.1103/PhysRevLett.124.121802}{\emph{Phys. Rev.
  Lett.} {\bf 124} (2020) 121802},
  [\href{https://arxiv.org/abs/1906.10745}{{\tt 1906.10745}}].

\bibitem{Miranda:2019skf}
O.~Miranda, G.~Sanchez~Garcia and O.~Sanders, \emph{{Coherent elastic
  neutrino-nucleus scattering as a precision test for the Standard Model and
  beyond: the COHERENT proposal case}},
  \href{http://dx.doi.org/10.1155/2019/3902819}{\emph{Adv. High Energy Phys.}
  {\bf 2019} (2019) 3902819}, [\href{https://arxiv.org/abs/1902.09036}{{\tt
  1902.09036}}].

\bibitem{Papoulias:2019txv}
D.~K. Papoulias, \emph{{COHERENT constraints after the Chicago-3 quenching
  factor measurement}},  \href{https://arxiv.org/abs/1907.11644}{{\tt
  1907.11644}}.

\bibitem{Khan:2019cvi}
A.~N. Khan and W.~Rodejohann, \emph{{New physics from COHERENT data with an
  improved quenching factor}},
  \href{http://dx.doi.org/10.1103/PhysRevD.100.113003}{\emph{Phys. Rev. D} {\bf
  100} (2019) 113003}, [\href{https://arxiv.org/abs/1907.12444}{{\tt
  1907.12444}}].

\bibitem{Cadeddu:2019eta}
M.~Cadeddu, F.~Dordei, C.~Giunti, Y.~Li and Y.~Zhang, \emph{{Neutrino,
  electroweak, and nuclear physics from COHERENT elastic neutrino-nucleus
  scattering with refined quenching factor}},
  \href{http://dx.doi.org/10.1103/PhysRevD.101.033004}{\emph{Phys. Rev. D} {\bf
  101} (2020) 033004}, [\href{https://arxiv.org/abs/1908.06045}{{\tt
  1908.06045}}].

\bibitem{Giunti:2019xpr}
C.~Giunti, \emph{{General COHERENT constraints on neutrino nonstandard
  interactions}},
  \href{http://dx.doi.org/10.1103/PhysRevD.101.035039}{\emph{Phys. Rev. D} {\bf
  101} (2020) 035039}, [\href{https://arxiv.org/abs/1909.00466}{{\tt
  1909.00466}}].

\bibitem{Han:2019zkz}
T.~Han, J.~Liao, H.~Liu and D.~Marfatia, \emph{{Nonstandard neutrino
  interactions at COHERENT, DUNE, T2HK and LHC}},
  \href{http://dx.doi.org/10.1007/JHEP11(2019)028}{\emph{JHEP} {\bf 11} (2019)
  028}, [\href{https://arxiv.org/abs/1910.03272}{{\tt 1910.03272}}].

\bibitem{AristizabalSierra:2019ykk}
D.~Aristizabal~Sierra, B.~Dutta, S.~Liao and L.~E. Strigari, \emph{{Coherent
  elastic neutrino-nucleus scattering in multi-ton scale dark matter
  experiments: Classification of vector and scalar interactions new physics
  signals}}, \href{http://dx.doi.org/10.1007/JHEP12(2019)124}{\emph{JHEP} {\bf
  12} (2019) 124}, [\href{https://arxiv.org/abs/1910.12437}{{\tt 1910.12437}}].

\bibitem{Papoulias:2019xaw}
D.~Papoulias, T.~Kosmas and Y.~Kuno, \emph{{Recent probes of standard and
  non-standard neutrino physics with nuclei}},
  \href{http://dx.doi.org/10.3389/fphy.2019.00191}{\emph{Front. in Phys.} {\bf
  7} (2019) 191}, [\href{https://arxiv.org/abs/1911.00916}{{\tt 1911.00916}}].

\bibitem{Arcadi:2019uif}
G.~Arcadi, M.~Lindner, J.~Martins and F.~S. Queiroz, \emph{{New Physics Probes:
  Atomic Parity Violation, Polarized Electron Scattering and Neutrino-Nucleus
  Coherent Scattering}},  \href{https://arxiv.org/abs/1906.04755}{{\tt
  1906.04755}}.

\bibitem{Miranda:2020zji}
O.~Miranda, D.~Papoulias, M.~Tórtola and J.~Valle, \emph{{Probing new neutral
  gauge bosons with $CE\nu NS$ and neutrino-electron scattering}},
  \href{http://dx.doi.org/10.1103/PhysRevD.101.073005}{\emph{Phys. Rev. D} {\bf
  101} (2020) 073005}, [\href{https://arxiv.org/abs/2002.01482}{{\tt
  2002.01482}}].

\bibitem{Sahu:2020kwh}
R.~Sahu, D.~Papoulias, V.~Kota and T.~Kosmas, \emph{{Elastic and inelastic
  scattering of neutrinos and WIMPs on nuclei}},
  \href{https://arxiv.org/abs/2004.04055}{{\tt 2004.04055}}.

\bibitem{Pattavina:2020cqc}
L.~Pattavina, N.~F. Iachellini and I.~Tamborra, \emph{{RES-NOVA: A
  revolutionary neutrino observatory based on archaeological lead}},
  \href{https://arxiv.org/abs/2004.06936}{{\tt 2004.06936}}.

\bibitem{Han:2020pff}
T.~Han, J.~Liao, H.~Liu and D.~Marfatia, \emph{{Scalar and tensor neutrino
  interactions}},  \href{https://arxiv.org/abs/2004.13869}{{\tt 2004.13869}}.

\bibitem{Foguel:2020fjx}
A.~L. Foguel, E.~S. Fraga and C.~Bonifazi, \emph{{Supernovae neutrino detection
  via coherent scattering off silicon nuclei}},
  \href{https://arxiv.org/abs/2005.13068}{{\tt 2005.13068}}.

\bibitem{Cadeddu:2020lky}
M.~Cadeddu, F.~Dordei, C.~Giunti, Y.~Li, E.~Picciau and Y.~Zhang,
  \emph{{Physics results from the first COHERENT observation of CE$\nu$NS in
  argon and their combination with cesium-iodide data}},
  \href{https://arxiv.org/abs/2005.01645}{{\tt 2005.01645}}.

\bibitem{Sadhukhan:2020etu}
S.~Sadhukhan and M.~P. Singh, \emph{{Neutrino Floor in Leptophilic $U(1)$
  Models: Modification in $U(1)_{L_{\mu}-L_{\tau}}$}},
  \href{https://arxiv.org/abs/2006.05981}{{\tt 2006.05981}}.

\bibitem{Coloma:2020nhf}
P.~Coloma, I.~Esteban, M.~Gonzalez-Garcia and J.~Menendez, \emph{{Determining
  the nuclear neutron distribution from Coherent Elastic neutrino-Nucleus
  Scattering: current results and future prospects}},
  \href{https://arxiv.org/abs/2006.08624}{{\tt 2006.08624}}.

\bibitem{Dutta:2020vop}
B.~Dutta, D.~Kim, S.~Liao, J.-C. Park, S.~Shin, L.~E. Strigari et~al.,
  \emph{{Searching for Dark Matter Signals in Timing Spectra at Neutrino
  Experiments}},  \href{https://arxiv.org/abs/2006.09386}{{\tt 2006.09386}}.

\bibitem{1805471}
N.~Van~Dessel, V.~Pandey, H.~Ray and N.~Jachowicz, \emph{{Nuclear Structure
  Physics in Coherent Elastic Neutrino-Nucleus Scattering}},
  \href{https://arxiv.org/abs/2007.03658}{{\tt 2007.03658}}.

\bibitem{Hoferichter:2020osn}
M.~Hoferichter, J.~Menéndez and A.~Schwenk, \emph{{Coherent elastic
  neutrino-nucleus scattering: EFT analysis and nuclear responses}},
  \href{https://arxiv.org/abs/2007.08529}{{\tt 2007.08529}}.

\bibitem{Skiba:2020msb}
W.~Skiba and Q.~Xia, \emph{{Electroweak Constraints from the COHERENT
  Experiment}},  \href{https://arxiv.org/abs/2007.15688}{{\tt 2007.15688}}.

\bibitem{Dutta:2020enk}
B.~Dutta, S.~Ghosh and J.~Kumar, \emph{{Opportunities for probing $U(1)_{T3R}$
  with light mediators}},  \href{https://arxiv.org/abs/2007.16191}{{\tt
  2007.16191}}.

\bibitem{1810387}
O.~Miranda, D.~Papoulias, O.~Sanders, M.~Tórtola and J.~Valle, \emph{{Future
  CEvNS experiments as probes of lepton unitarity and light-sterile
  neutrinos}},  \href{https://arxiv.org/abs/2008.02759}{{\tt 2008.02759}}.

\bibitem{Cadeddu:2020nbr}
M.~Cadeddu, N.~Cargioli, F.~Dordei, C.~Giunti, Y.~Li, E.~Picciau et~al.,
  \emph{{Constraints on light vector mediators through coherent elastic
  neutrino nucleus scattering data from COHERENT}},
  \href{https://arxiv.org/abs/2008.05022}{{\tt 2008.05022}}.

\bibitem{Tomalak:2020zfh}
O.~Tomalak, P.~Machado, V.~Pandey and R.~Plestid, \emph{{Flavor-dependent
  radiative corrections in coherent elastic neutrino-nucleus scattering}},
  \href{https://arxiv.org/abs/2011.05960}{{\tt 2011.05960}}.

\bibitem{Suliga:2020jfa}
A.~M. Suliga and I.~Tamborra, \emph{{Astrophysical constraints on non-standard
  coherent neutrino-nucleus scattering}},
  \href{https://arxiv.org/abs/2010.14545}{{\tt 2010.14545}}.

\bibitem{Akimov:2018vzs}
{\scshape COHERENT} collaboration, D.~Akimov et~al., \emph{{COHERENT
  Collaboration data release from the first observation of coherent elastic
  neutrino-nucleus scattering}},  \href{https://arxiv.org/abs/1804.09459}{{\tt
  1804.09459}}.

\bibitem{Feldman:1997qc}
G.~J. Feldman and R.~D. Cousins, \emph{{A Unified approach to the classical
  statistical analysis of small signals}},
  \href{http://dx.doi.org/10.1103/PhysRevD.57.3873}{\emph{Phys. Rev. D} {\bf
  57} (1998) 3873--3889}, [\href{https://arxiv.org/abs/physics/9711021}{{\tt
  physics/9711021}}].

\bibitem{7stats}
J.~Gehrlein, ``7stats.'' \url{https://github.com/JuliaGehrlein/7stats}.

\bibitem{Barranco:2005yy}
J.~Barranco, O.~Miranda and T.~Rashba, \emph{{Probing new physics with coherent
  neutrino scattering off nuclei}},
  \href{http://dx.doi.org/10.1088/1126-6708/2005/12/021}{\emph{JHEP} {\bf 12}
  (2005) 021}, [\href{https://arxiv.org/abs/hep-ph/0508299}{{\tt
  hep-ph/0508299}}].

\bibitem{Beacom:2002hs}
J.~F. Beacom, W.~M. Farr and P.~Vogel, \emph{{Detection of Supernova Neutrinos
  by Neutrino Proton Elastic Scattering}},
  \href{http://dx.doi.org/10.1103/PhysRevD.66.033001}{\emph{Phys.\ Rev.\ D}
  {\bf 66} (2002) 033001}, [\href{https://arxiv.org/abs/hep-ph/0205220}{{\tt
  hep-ph/0205220}}].

\bibitem{Klein:1999qj}
S.~Klein and J.~Nystrand, \emph{{Exclusive vector meson production in
  relativistic heavy ion collisions}},
  \href{http://dx.doi.org/10.1103/PhysRevC.60.014903}{\emph{Phys.\ Rev.\ C}
  {\bf 60} (1999) 014903}, [\href{https://arxiv.org/abs/hep-ph/9902259}{{\tt
  hep-ph/9902259}}].

\bibitem{Coloma:2019mbs}
P.~Coloma, I.~Esteban, M.~Gonzalez-Garcia and M.~Maltoni, \emph{{Improved
  global fit to Non-Standard neutrino Interactions using COHERENT energy and
  timing data}}, \href{http://dx.doi.org/10.1007/JHEP02(2020)023}{\emph{JHEP}
  {\bf 02} (2020) 023}, [\href{https://arxiv.org/abs/1911.09109}{{\tt
  1911.09109}}].

\bibitem{Wolfenstein:1977ue}
L.~Wolfenstein, \emph{{Neutrino Oscillations in Matter}},
  \href{http://dx.doi.org/10.1103/PhysRevD.17.2369}{\emph{Phys.\ Rev.\ D} {\bf
  17} (1978) 2369--2374}.

\bibitem{Forero:2016ghr}
D.~V. Forero and W.-C. Huang, \emph{{Sizable NSI from the SU(2)$_{L}$ scalar
  doublet-singlet mixing and the implications in DUNE}},
  \href{http://dx.doi.org/10.1007/JHEP03(2017)018}{\emph{JHEP} {\bf 03} (2017)
  018}, [\href{https://arxiv.org/abs/1608.04719}{{\tt 1608.04719}}].

\bibitem{Denton:2018dqq}
P.~B. Denton, Y.~Farzan and I.~M. Shoemaker, \emph{{Activating the fourth
  neutrino of the 3+1 scheme}},
  \href{http://dx.doi.org/10.1103/PhysRevD.99.035003}{\emph{Phys. Rev. D} {\bf
  99} (2019) 035003}, [\href{https://arxiv.org/abs/1811.01310}{{\tt
  1811.01310}}].

\bibitem{Dey:2018yht}
U.~K. Dey, N.~Nath and S.~Sadhukhan, \emph{{Non-Standard Neutrino Interactions
  in a Modified $\nu$2HDM}},
  \href{http://dx.doi.org/10.1103/PhysRevD.98.055004}{\emph{Phys. Rev. D} {\bf
  98} (2018) 055004}, [\href{https://arxiv.org/abs/1804.05808}{{\tt
  1804.05808}}].

\bibitem{Babu:2017olk}
K.~Babu, A.~Friedland, P.~Machado and I.~Mocioiu, \emph{{Flavor Gauge Models
  Below the Fermi Scale}},
  \href{http://dx.doi.org/10.1007/JHEP12(2017)096}{\emph{JHEP} {\bf 12} (2017)
  096}, [\href{https://arxiv.org/abs/1705.01822}{{\tt 1705.01822}}].

\bibitem{Farzan:2016wym}
Y.~Farzan and J.~Heeck, \emph{{Neutrinophilic nonstandard interactions}},
  \href{http://dx.doi.org/10.1103/PhysRevD.94.053010}{\emph{Phys. Rev. D} {\bf
  94} (2016) 053010}, [\href{https://arxiv.org/abs/1607.07616}{{\tt
  1607.07616}}].

\bibitem{Farzan:2015hkd}
Y.~Farzan and I.~M. Shoemaker, \emph{{Lepton Flavor Violating Non-Standard
  Interactions via Light Mediators}},
  \href{http://dx.doi.org/10.1007/JHEP07(2016)033}{\emph{JHEP} {\bf 07} (2016)
  033}, [\href{https://arxiv.org/abs/1512.09147}{{\tt 1512.09147}}].

\bibitem{Farzan:2015doa}
Y.~Farzan, \emph{{A model for large non-standard interactions of neutrinos
  leading to the LMA-Dark solution}},
  \href{http://dx.doi.org/10.1016/j.physletb.2015.07.015}{\emph{Phys. Lett. B}
  {\bf 748} (2015) 311--315}, [\href{https://arxiv.org/abs/1505.06906}{{\tt
  1505.06906}}].

\bibitem{Babu:2019mfe}
K.~Babu, P.~B. Dev, S.~Jana and A.~Thapa, \emph{{Non-Standard Interactions in
  Radiative Neutrino Mass Models}},
  \href{http://dx.doi.org/10.1007/JHEP03(2020)006}{\emph{JHEP} {\bf 03} (2020)
  006}, [\href{https://arxiv.org/abs/1907.09498}{{\tt 1907.09498}}].

\bibitem{Girardi:2014kca}
I.~Girardi, D.~Meloni and S.~Petcov, \emph{{The Daya Bay and T2K results on
  $\sin^2 2 \theta_{13}$ and Non-Standard Neutrino Interactions}},
  \href{http://dx.doi.org/10.1016/j.nuclphysb.2014.06.014}{\emph{Nucl. Phys. B}
  {\bf 886} (2014) 31--42}, [\href{https://arxiv.org/abs/1405.0416}{{\tt
  1405.0416}}].

\bibitem{Liao:2016reh}
J.~Liao and D.~Marfatia, \emph{{Impact of nonstandard interactions on sterile
  neutrino searches at IceCube}},
  \href{http://dx.doi.org/10.1103/PhysRevLett.117.071802}{\emph{Phys. Rev.
  Lett.} {\bf 117} (2016) 071802},
  [\href{https://arxiv.org/abs/1602.08766}{{\tt 1602.08766}}].

\bibitem{Capozzi:2019iqn}
F.~Capozzi, S.~S. Chatterjee and A.~Palazzo, \emph{{Neutrino Mass Ordering
  Obscured by Nonstandard Interactions}},
  \href{http://dx.doi.org/10.1103/PhysRevLett.124.111801}{\emph{Phys. Rev.
  Lett.} {\bf 124} (2020) 111801},
  [\href{https://arxiv.org/abs/1908.06992}{{\tt 1908.06992}}].

\bibitem{Denton:2020uda}
P.~B. Denton, J.~Gehrlein and R.~Pestes, \emph{{$CP$ -Violating Neutrino
  Nonstandard Interactions in Long-Baseline-Accelerator Data}},
  \href{http://dx.doi.org/10.1103/PhysRevLett.126.051801}{\emph{Phys. Rev.
  Lett.} {\bf 126} (2021) 051801},
  [\href{https://arxiv.org/abs/2008.01110}{{\tt 2008.01110}}].

\bibitem{Dev:2019anc}
P.~Bhupal~Dev et~al., \emph{{Neutrino Non-Standard Interactions: A Status
  Report}},
  \href{http://dx.doi.org/10.21468/SciPostPhysProc.2.001}{\emph{SciPost Phys.\
  Proc.} {\bf 2} (2019) 001}, [\href{https://arxiv.org/abs/1907.00991}{{\tt
  1907.00991}}].

\bibitem{Collar:2019ihs}
J.~Collar, A.~Kavner and C.~Lewis, \emph{{Response of CsI[Na] to Nuclear
  Recoils: Impact on Coherent Elastic Neutrino-Nucleus Scattering
  (CE$\nu$NS)}},
  \href{http://dx.doi.org/10.1103/PhysRevD.100.033003}{\emph{Phys. Rev. D} {\bf
  100} (2019) 033003}, [\href{https://arxiv.org/abs/1907.04828}{{\tt
  1907.04828}}].

\bibitem{Konovalov2019}
A.~Konovalov, ``{CsI[Na] effort of the COHERENT collaboration}.''
  \href{https://indico.cern.ch/event/844613/contributions/3615323/attachments/1942027/3220535/Mag_CEvNS_Konovalov.pdf}{Talk
  at Magnificent CEvNS 2019}, 2019.

\bibitem{Rich:2017lzd}
G.~C. Rich, \emph{{Measurement of Low-Energy Nuclear-Recoil Quenching Factors
  in CsI[Na] and Statistical Analysis of the First Observation of Coherent,
  Elastic Neutrino-Nucleus Scattering}}.
\newblock PhD thesis, North Carolina U., 2017.

\bibitem{Scholz:2017ldm}
B.~J. Scholz, \emph{{First Observation of Coherent Elastic Neutrino-Nucleus
  Scattering}}.
\newblock PhD thesis, Chicago U., 2017.
\newblock \href{https://arxiv.org/abs/1904.01155}{{\tt 1904.01155}}.
\newblock 10.1007/978-3-319-99747-6.

\bibitem{Guo:2016ovb}
C.~Guo et~al., \emph{{Neutron beam tests of $CsI(Na)$ and $CaF_{2}(Eu)$
  crystals for dark matter direct search}},
  \href{http://dx.doi.org/10.1016/j.nima.2016.02.037}{\emph{Nucl. Instrum.
  Meth. A} {\bf 818} (2016) 38--44},
  [\href{https://arxiv.org/abs/1602.04923}{{\tt 1602.04923}}].

\bibitem{Park:2002jr}
H.~Park et~al., \emph{{Neutron beam test of CsI crystal for dark matter
  search}}, \href{http://dx.doi.org/10.1016/S0168-9002(02)01274-3}{\emph{Nucl.
  Instrum. Meth. A} {\bf 491} (2002) 460--469},
  [\href{https://arxiv.org/abs/nucl-ex/0202014}{{\tt nucl-ex/0202014}}].

\bibitem{Tolstukhin2019}
I.~Tolstukhin, ``{COHERENT experiment at the Spallation Neutron Source}.''
  \href{https://absuploads.aps.org/presentation.cfm?pid=14852}{Talk at APS
  April Meeting 2019}, 2019.

\bibitem{Demortier}
L.~Demortier and L.~Lyons, ``{Everything you always wanted to know about
  pulls}.''
  \href{http://physics.rockefeller.edu/luc/technical_reports/cdf5776_pulls.pdf}{CDF/ANAL/PUBLIC/5776},
  2002.

\bibitem{NOVA-doc-15884-v3}
G.~Feldman, ``Notes on the inclusion of nuisance parameters in the unified
  approach.''
  \href{https://nova-docdb.fnal.gov/cgi-bin/ShowDocument?docid=15884}{NOVA-doc-15884},
  2016.

\bibitem{10.2307/2290928}
R.~L. Berger and D.~D. Boos, \emph{P values maximized over a confidence set for
  the nuisance parameter}, {\emph{Journal of the American Statistical
  Association} {\bf 89} (1994) 1012--1016}.

\bibitem{Wilks:1938dza}
S.~Wilks, \emph{{The Large-Sample Distribution of the Likelihood Ratio for
  Testing Composite Hypotheses}},
  \href{http://dx.doi.org/10.1214/aoms/1177732360}{\emph{Annals Math. Statist.}
  {\bf 9} (1938) 60--62}.

\bibitem{Lyons:2014kta}
L.~Lyons, \emph{{Raster scan or 2-D approach?}},
  \href{https://arxiv.org/abs/1404.7395}{{\tt 1404.7395}}.

\bibitem{Algeri:2019arh}
S.~Algeri, J.~Aalbers, K.~Dundas~Mora and J.~Conrad, \emph{{Searching for new
  physics with profile likelihoods: Wilks and beyond}},
  \href{https://arxiv.org/abs/1911.10237}{{\tt 1911.10237}}.

\bibitem{Agostini:2019jup}
M.~Agostini and B.~Neumair, \emph{{Statistical Methods for the Search of
  Sterile Neutrinos}},  \href{https://arxiv.org/abs/1906.11854}{{\tt
  1906.11854}}.

\bibitem{Silaeva:2020yot}
S.~Silaeva and V.~Sinev, \emph{{Simulation of an experiment on looking for
  sterile neutrinos at nuclear reactor}},
  \href{https://arxiv.org/abs/2001.10752}{{\tt 2001.10752}}.

\bibitem{Giunti:2020uhv}
C.~Giunti, \emph{{Statistical Significance of Reactor Antineutrino
  Active-Sterile Oscillations}},  \href{https://arxiv.org/abs/2004.07577}{{\tt
  2004.07577}}.

\bibitem{Berryman:2020agd}
J.~M. Berryman and P.~Huber, \emph{{Sterile Neutrinos and the Global Reactor
  Antineutrino Dataset}},  \href{https://arxiv.org/abs/2005.01756}{{\tt
  2005.01756}}.

\bibitem{Almazan:2020drb}
H.~Almaza'n et~al., \emph{{Note on arXiv:2005.05301, 'Preparation of the
  Neutrino-4 experiment on search for sterile neutrino and the obtained results
  of measurements'}},  \href{https://arxiv.org/abs/2006.13147}{{\tt
  2006.13147}}.

\bibitem{Qian:2014nha}
X.~Qian, A.~Tan, J.~J. Ling, Y.~Nakajima and C.~Zhang, \emph{{The Gaussian
  CL$_s$ method for searches of new physics}},
  \href{http://dx.doi.org/10.1016/j.nima.2016.04.089}{\emph{Nucl. Instrum.
  Meth. A} {\bf 827} (2016) 63--78},
  [\href{https://arxiv.org/abs/1407.5052}{{\tt 1407.5052}}].

\bibitem{Coloma:2020ajw}
P.~Coloma, P.~Huber and T.~Schwetz, \emph{{Statistical interpretation of
  sterile neutrino oscillation searches at reactors}},
  \href{https://arxiv.org/abs/2008.06083}{{\tt 2008.06083}}.

\bibitem{Blennow:2014sja}
M.~Blennow, P.~Coloma and E.~Fernandez-Martinez, \emph{{Reassessing the
  sensitivity to leptonic CP violation}},
  \href{http://dx.doi.org/10.1007/JHEP03(2015)005}{\emph{JHEP} {\bf 03} (2015)
  005}, [\href{https://arxiv.org/abs/1407.3274}{{\tt 1407.3274}}].

\bibitem{Blennow:2013oma}
M.~Blennow, P.~Coloma, P.~Huber and T.~Schwetz, \emph{{Quantifying the
  sensitivity of oscillation experiments to the neutrino mass ordering}},
  \href{http://dx.doi.org/10.1007/JHEP03(2014)028}{\emph{JHEP} {\bf 03} (2014)
  028}, [\href{https://arxiv.org/abs/1311.1822}{{\tt 1311.1822}}].

\bibitem{Schwetz:2006md}
T.~Schwetz, \emph{{What is the probability that theta(13) and CP violation will
  be discovered in future neutrino oscillation experiments?}},
  \href{http://dx.doi.org/10.1016/j.physletb.2007.02.053}{\emph{Phys. Lett. B}
  {\bf 648} (2007) 54--59}, [\href{https://arxiv.org/abs/hep-ph/0612223}{{\tt
  hep-ph/0612223}}].

\bibitem{Elevant:2015ska}
J.~Elevant and T.~Schwetz, \emph{{On the determination of the leptonic CP
  phase}}, \href{http://dx.doi.org/10.1007/JHEP09(2015)016}{\emph{JHEP} {\bf
  09} (2015) 016}, [\href{https://arxiv.org/abs/1506.07685}{{\tt 1506.07685}}].

\bibitem{Akimov:2020pdx}
{\scshape COHERENT} collaboration, D.~Akimov et~al., \emph{{First Detection of
  Coherent Elastic Neutrino-Nucleus Scattering on Argon}},
  \href{https://arxiv.org/abs/2003.10630}{{\tt 2003.10630}}.

\bibitem{Daughhetee2020}
J.~Daughhetee, ``{Detection of CEvNS on Argon in the CENNS-10 Liquid Argon
  Detector}.''
  \href{https://absuploads.aps.org/presentation.cfm?pid=18388}{Talk at APS
  April Meeting 2020}, 2020.

\bibitem{Pershey:2018gtf}
D.~S. Pershey, \emph{{A Measurement of $\nu_e$ Appearance and $\nu_\mu$
  Disappearance Neutrino Oscillations with the NOvA Experiment}}.
\newblock PhD thesis, Caltech, 2018.
\newblock 10.2172/1484186.

\bibitem{Esteban:2018ppq}
I.~Esteban, M.~Gonzalez-Garcia, M.~Maltoni, I.~Martinez-Soler and J.~Salvado,
  \emph{{Updated Constraints on Non-Standard Interactions from Global Analysis
  of Oscillation Data}},
  \href{http://dx.doi.org/10.1007/JHEP08(2018)180}{\emph{JHEP} {\bf 08} (2018)
  180}, [\href{https://arxiv.org/abs/1805.04530}{{\tt 1805.04530}}].

\end{thebibliography}\endgroup

\end{document}